\documentclass[12pt]{article}

\usepackage{amsfonts,amssymb,graphics,epsfig,verbatim,bm,latexsym,amsmath,url,amsbsy,
geometry}
\usepackage{longtable}
\usepackage{amsthm}
\newtheorem{theorem}{Theorem}

\newtheorem{lemma}{Lemma}
\newtheorem*{lemma*}{Lemma A.1}
\usepackage{csquotes}
\usepackage[authoryear]{natbib}
\usepackage{setspace}
\usepackage{sectsty}
\usepackage[tableposition=top]{caption}
\sectionfont{\centering}

\newtheorem{innercustomgeneric}{\customgenericname}
\providecommand{\customgenericname}{}
\newcommand{\newcustomtheorem}[2]{%
  \newenvironment{#1}[1]
  {%
   \renewcommand\customgenericname{#2}%
   \renewcommand\theinnercustomgeneric{##1}%
   \innercustomgeneric
  }
  {\endinnercustomgeneric}
}

\newcustomtheorem{customthm}{Theorem}
\newcustomtheorem{customlemma}{Lemma}

\bibpunct{(}{)}{;}{a}{,}{,}

 \usepackage[utf8]{inputenc}
\usepackage{pgfplots}
\pgfplotsset{compat=1.9}

\begin{document}
\bibliographystyle{plainnat}
\onehalfspacing

\title{\bf Forward Orthogonal Deviations GMM and the Absence of Large Sample Bias}
\date{July 2024}
\author{Robert F. Phillips\footnote{2115 G Street, NW, Suite 340, Washington DC, 20052; phone: 202-994-8619; fax: 202-994-6147; email: \texttt{rphil@gwu.edu}} \\Department of Economics\\George Washington University}
\begin{titlepage}
\clearpage\maketitle
\thispagestyle{empty}

 \abstract{It is well known that generalized method of moments (GMM) estimators of dynamic panel data regressions  can have significant bias when the number of time periods ($T$) is not small compared to the number of cross-sectional units ($n$).  The bias is attributed to the use of many instrumental variables. This paper shows that if the maximum number of instrumental variables used in a period increases with $T$ at a rate slower than $T^{1/2}$, then GMM estimators that exploit the forward orthogonal deviations (FOD) transformation  do not have asymptotic bias,  regardless of how fast $T$ increases relative to $n$.  This conclusion is specific to using the FOD transformation.  A similar conclusion does not necessarily apply when other transformations are used to remove fixed effects.    Monte Carlo evidence illustrating the analytical results is provided. \\}
 
\end{titlepage}
\newpage

\section{Introduction \label{intro}}

In an influential paper, \cite{Alvarez2003} examined the asymptotic properties of a generalized method of moments (GMM) estimator that relied on  forward orthogonal deviations (FOD) to remove fixed effects (FOD GMM).   They showed that an FOD GMM estimator of the autoregression parameter in a first-order autoregressive (AR(1)) panel data model has a bias term in its asymptotic distribution if the number of time periods ($T$) increases too quickly relative to the number of cross-sectional units ($n$)---that is, if $T/n$ does not converge to zero.  Though the model considered by \cite{Alvarez2003} was a special case, the paper's finding had an important implication.  This is because if an estimator, say $\widehat{\delta}$, of parameter $\delta$ has a bias term in the asymptotic distribution of $\sqrt{nT}(\widehat{\delta} - \delta)$, then even though the estimator may be consistent,  large sample confidence intervals and test statistics based on the estimator will be inaccurate.  

Alvarez and Arellano's conclusion that the FOD GMM estimator they studied has asymptotic bias if $T/n$ does not converge to zero depends on  using all available instrumental variables.  If $T$ is not small,  the number of instrumental variables can be large if all of them are used, and it is well-known that a GMM estimator that exploits many instrumental variables can be biased.  Consequently, in practice researchers often resort to using fewer instrumental variables than all that are available.  But left unanswered is the question: how many instrumental variables can be used while still avoiding bias when $T$ is not small compared to $n$?  This paper addresses that question.  

\sloppy
Like this paper, \cite{Anderson2011}, \cite{Bekker1994}, \cite{Bun2006}, \cite{Chao2005},  \cite{Hansen2008}, \cite{Hayakawa2019}, and \cite{Koenker1999}, among others, study how the properties of estimators are affected by the number of moment restrictions that are exploited. However, the papers most closely related to this paper are \cite{Alvarez2003} and \cite{Hsiao2017}.  
  As already noted, \cite{Alvarez2003} considered estimation of the AR(1) panel data model:
\begin{equation} \label{ar1_model}
	y_{i,t} = \beta y_{i,t-1} + \eta_{i} + v_{i,t}, \qquad \left|\beta\right| < 1.
\end{equation}
Assuming all available moment restrictions are exploited, Alvarez and Arelleno found that an FOD GMM estimator, say $\widehat{\beta}$, of the autoregression parameter, $\beta$, has a bias term in the asymptotic distribution of $\sqrt{nT}(\widehat{\beta} - \beta)$ if $T/n \rightarrow c >0$, as $n,T \rightarrow \infty$, but it does not have a bias term if $T/n \rightarrow 0$.   \cite{Hsiao2017}, on the other hand,  investigated the effect of using fewer than all available moment restrictions.  In particular, \cite{Hsiao2017}  analyzed FOD GMM estimation of the  model in (\ref{ar1_model}), but, instead of focusing on only using all available instrumental variables, Hsiao and Zhou also considered estimation based on a single instrumental variable  per period.  Upon doing so, they obtained results indicating that the FOD GMM estimators they considered---when based on a single instrumental variable per period---have no bias in their asymptotic distributions, as $n,T \rightarrow \infty$, regardless of what happens to $T/n$.\footnote{On the other hand, \cite{Hsiao2017} also argued that GMM based on first differences (FD GMM) and a single instrumental variable per period is asymptotically biased if $T/n \rightarrow c > 0$.} The papers by \cite{Alvarez2003} and \cite{Hsiao2017}, therefore, indicate that whether or not there is a bias term in the asymptotic distribution of an FOD GMM estimator depends not just on what happens to $T/n$, as $n, T \rightarrow \infty$, but also on how many per-period instruments are used.

\fussy
Using more general conditions than previously considered, this paper shows that whether or not an FOD GMM estimator is asymptotically biased can be summarized in terms of the largest number of instrumental variables used in a  period.  Specifically,  I show  that the FOD GMM estimator has no asymptotic bias regardless of what happens to $T/n$,  if the maximum number of instruments used in a period increases with $T$ at a rate slower than $T^{1/2}$ increases. This conclusion is specific to using the FOD transformation.  It does not necessarily carry over to other transformations that may be used to remove fixed effects.  

Moreover, the conclusion is robust in the sense that the  large sample bias result holds regardless of how $n$ and $T$ increase.  Specifically, the conclusion is based on taking joint limits, which are limits obtained by letting $n$ and $T$ increase simultaneously.  

A second less robust result is also provided: an asymptotic distribution result is provided that is based on taking limits sequentially.  With a sequential limit,  one index---$n$ or $T$---is taken to infinity, and then the other goes to infinity. Taking limits sequentially is often a more tractable tactic for obtaining results than joint limits, and sequential limit results may require weaker conditions than joint limit results.  These advantages may explain why limits are sometimes taken sequentially  in the literature. \cite{Hsiao2017}, for example, used sequential limits to derive asymptotic distribution results for their FOD GMM estimators of the autoregression parameter in the model in (\ref{ar1_model}).  A few other papers that exploit sequential limits are \cite{Kapetanios2008} and \cite{Hsiao2015}.  

However, sequential limits are not guaranteed to yield the same results one gets from joint limit analysis \citep{Moon1999, Moon2000}.  Indeed, I show that this is the case when all available instrumental variables are used.  On the other hand, when fewer than all available instruments are used, Monte Carlo evidence indicates that the normal approximation obtained by taking limits sequentially appears to work well for constructing confidence intervals when $T$---in addition to $n$---is not small, provided the FOD transformation is used to remove fixed effects.

\section{FOD GMM \label{analysis}}
\subsection{The model and estimator}

The regression model studied in this paper is
\begin{equation*} \label{model}
	y_{i,t} = \boldsymbol{x}_{i,t}^{\prime}\boldsymbol{\beta} + \eta_{i} + v_{i,t} \qquad (t=1,\ldots,T, \; i=1,\ldots,n).
\end{equation*}
In this regression, $\boldsymbol{x}_{i,t}' := (x_{i,t,1},\ldots, x_{i,t,K})$ and $\boldsymbol{\beta}' := (\beta_1, \ldots, \beta_K)$  are vectors of regressors and parameters. Some  or all of the $x_{i,t,k}$s may be lagged values of $ y_{i,t} $. The term $ \eta_i $ is an unobserved individual-specific  or fixed effect, and  $ v_{i,t} $ is an error term.  

In order to estimate $\boldsymbol{\beta}$, the first step is to remove the fixed effect $\eta_i$ by transforming the dependent and explanatory variables. This paper studies transforming the variables using forward orthogonal deviations---the FOD transformation. The FOD transformation subtracts from each variable its within-group average over future periods. For example, the FOD transformed explanatory variables for the $i$th individual in the $t$th period are  
$ \ddot{\boldsymbol{x}}_{i,t} := c_{t} \left(   \boldsymbol{x}_{i,t} - \overline{\boldsymbol{x}}_{i,t} \right) $, 
where $ \overline{\boldsymbol{x}}_{i,t} := \left( 1/(T-t)\right)\sum_{s=1}^{T-t} \boldsymbol{x}_{i,t+s} $ and  $c_t^2 := (T-t)/(T-t+1) $.  Similarly, the transformed value of the ($i,t$)th observation on the dependent variable is $ \ddot{y}_{i,t} := c_{t} \left(   y_{i,t} - \overline{y}_{i,t} \right) $, 
where  $ \overline{y}_{i,t} := \left( 1/(T-t)\right)\sum_{s=1}^{T-t} y_{i,t+s} $.  The constant $c_t $ ensures the transformed errors (the $ \ddot{v}_{i,t} $s) are conditionally homoskedastic and uncorrelated if the original errors (the $ v_{i,t} $s) are conditionally homoskedastic and  uncorrelated \citep[p. 17]{Arellano2003}.  

Now let $ \boldsymbol{z}_{i,t} $ denote a $ q_{t} \times 1 $  vector of instrumental variables for the $ i $th individual in the $ t $th period ($t=1,\ldots,T-1$).  Also, set $ \boldsymbol{Z}_t' := \left(\boldsymbol{z}_{1,t}, \ldots, \boldsymbol{z}_{n,t} \right) $, $\ddot{\boldsymbol{X}}_t' := (\ddot{\boldsymbol{x}}_{1,t}, \ldots, \ddot{\boldsymbol{x}}_{n,t})$, $\ddot{\boldsymbol{y}}_t' := (\ddot{y}_{1,t}, \ldots, \ddot{y}_{n,t})$, and finally $ \boldsymbol{P}_t := \boldsymbol{Z}_t\left( \boldsymbol{Z}_t'\boldsymbol{Z}_t\right)^{-1}\boldsymbol{Z}_t'  $.  Then, the FOD GMM estimator can be written as
\begin{equation} \label{fod_est}
\widehat{\boldsymbol{\beta}} := \left( \sum_{t=1}^{T-1}\ddot{\boldsymbol{X}}_t'\boldsymbol{P}_t \ddot{\boldsymbol{X}}_t \right)^{-1}\,\sum_{t=1}^{T-1}\ddot{\boldsymbol{X}}_t'\boldsymbol{P}_t \ddot{\boldsymbol{y}}_t
\end{equation}
\citep[see, e.g.,][p. 154]{Arellano2003}.

\sloppy
As an alternative to Eq. (\ref{fod_est}), the FOD GMM estimator can also be expressed as a two-stage least squares (TSLS) estimator after removing fixed effects with the FOD transformation.  To see this, let $\boldsymbol{Z}_{d,i}$ be the block-diagonal matrix given by 
\begin{equation} \label{Z_di}
\boldsymbol{Z}_{d,i} := \left(
\begin{array}{cccc}
 \boldsymbol{z}_{i,1}' & \boldsymbol{0}  & \cdots & \boldsymbol{0} \\
\boldsymbol{0}  &  \boldsymbol{z}_{i,2}' & \cdots & \boldsymbol{0}\\
\vdots & \vdots & \ddots & \vdots\\
 \boldsymbol{0} &  \boldsymbol{0} & \cdots & \boldsymbol{z}_{i,T-1}'
\end{array}
\right).
\end{equation}
Also, define $\dot{\boldsymbol{X}}_i' := (\ddot{\boldsymbol{x}}_{i,1}, \ldots, \ddot{\boldsymbol{x}}_{i,T-1})$ and $\dot{\boldsymbol{y}}_i' := (\ddot{y}_{i,1}, \ldots, \ddot{y}_{i,T-1})$.  Then, note that $\sum_{t=1}^{T-1}\ddot{\boldsymbol{X}}_t'\boldsymbol{P}_t \ddot{\boldsymbol{X}}_t =  \sum_{i=1}^n \dot{\boldsymbol{X}}_i'\boldsymbol{Z}_{d,i}(\sum_{i=1}^n\boldsymbol{Z}_{d,i}'\boldsymbol{Z}_{d,i})^{-1}\sum_{i=1}^n\boldsymbol{Z}_{d,i}'\dot{\boldsymbol{X}}_i $. Similarly,  $\sum_{t=1}^{T-1}\ddot{\boldsymbol{X}}_t'\boldsymbol{P}_t \ddot{\boldsymbol{y}}_t =  \sum_{i=1}^n \dot{\boldsymbol{X}}_i'\boldsymbol{Z}_{d,i}(\sum_{i=1}^n\boldsymbol{Z}_{d,i}'\boldsymbol{Z}_{d,i})^{-1}\sum_{i=1}^n\boldsymbol{Z}_{d,i}'\dot{\boldsymbol{y}}_i $.  From this and Eq. (\ref{fod_est}), we see that the FOD GMM estimator can be expressed as a TSLS estimator after applying the FOD transformation to the dependent and explanatory variables and upon using block-diagonal instrument matrices:
\begin{equation*} \label{tsls}
\begin{split}
\widehat{\boldsymbol{\beta}}  = & \left[  \sum_{i=1}^n \dot{\boldsymbol{X}}_i'\boldsymbol{Z}_{d,i}\left(\sum_{i=1}^n\boldsymbol{Z}_{d,i}'\boldsymbol{Z}_{d,i}\right)^{-1}\sum_{i=1}^n\boldsymbol{Z}_{d,i}'\dot{\boldsymbol{X}}_i \right]^{-1} \\
& \times \sum_{i=1}^n \dot{\boldsymbol{X}}_i'\boldsymbol{Z}_{d,i}\left(\sum_{i=1}^n\boldsymbol{Z}_{d,i}'\boldsymbol{Z}_{d,i}\right)^{-1}\sum_{i=1}^n\boldsymbol{Z}_{d,i}'\dot{\boldsymbol{y}}_i .
\end{split}
\end{equation*}

\fussy
It is well-known that TSLS is efficient GMM when the errors are conditionally homoskedastic and uncorrelated.  Moreover, as already noted, if the $v_{i,t}$s are conditionally homoskedastic and uncorrelated, then the transformed errors (the $\ddot{v}_{i,t}$s) are conditionally homoskedastic and uncorrelated.   Hence, if the $v_{i,t}$s are conditionally homoskedastic and uncorrelated, the  FOD GMM estimator is the asymptotically efficient GMM estimator given the moment restrictions $E(\boldsymbol{z}_{i,t}\ddot{v}_{i,t}) = \boldsymbol{0}$ ($t=1,\ldots,T-1$).  This is a total of $\sum_{t=1}^{T-1}q_t$ moment restrictions, which can be a large number when $T$ is large, especially if $	q_{T}^{\ast} := \max_{1 \le t \le T-1}q_{t}$---i.e., the maximum per-period number of instrumental variables---increases with $T$.  Therefore, although the FOD GMM estimator is efficient when the errors are conditionally homoskedastic and uncorrelated, we might anticipate it to be biased when $T$ is not small.

\subsection{When there is no asymptotic bias \label{bias_analysis}}

\sloppy
\cite{Alvarez2003} provide conditions under which the distribution of an FOD GMM estimator, $\widehat{\beta}$, of the autoregression parameter, $\beta$, in the model in (\ref{ar1_model}) has an asymptotic bias term if  $T/n \rightarrow c > 0$, as $n,T \rightarrow \infty$.  In particular, note that
\begin{equation} \label{bias_ar1}
\sqrt{nT}\left(\widehat{\beta} - \beta\right) - \theta_{n,T}  = \frac{b_{n,T}- E(b_{n,T})}{a_{n,T}}, 
\end{equation}
where  $\theta_{n,T} := E(b_{n,T})/a_{n,T} $, $a_{n,T} := (1/(nT))\sum_{t=1}^{T-1}\ddot{\boldsymbol{y}}_{t-1}'\boldsymbol{P}_t \ddot{\boldsymbol{y}}_{t-1}$, $
b_{n,T} := (1/\sqrt{nT})\sum_{t=1}^{T-1}\ddot{\boldsymbol{y}}_{t-1}'\boldsymbol{P}_t \ddot{\boldsymbol{v}}_t $,  $\ddot{\boldsymbol{y}}_{t-1} := c_t(\boldsymbol{y}_{t-1} - \overline{\boldsymbol{y}}_{t-1})$, $\boldsymbol{y}_{t-1}' := (y_{1,t-1},\ldots,y_{n,t-1})$, and $\overline{\boldsymbol{y}}_{t-1} := \left( 1/(T-t)\right)\sum_{s=1}^{T-t} \boldsymbol{y}_{t-1+s}$. If $a_{n,T}$ converges in probability to a fixed non-zero limit and $b_{n,T} - E(b_{n,T})$ has a limit distribution,  then $\sqrt{nT}(\widehat{\beta} - \beta)$ has an asymptotic distribution that is centered at zero only if $\theta_{n,T}  \overset{p}\rightarrow 0$, as $n,T \rightarrow \infty$.  \cite{Alvarez2003} provide conditions that imply
\begin{equation*}
a_{n,T} \overset{p}{\rightarrow} \frac{\sigma^2}{1 - \beta^2} \qquad \text{and} \qquad E(b_{n,T}) - \sqrt{\frac{T}{n}}\left(\frac{\sigma^2}{ \beta - 1} \right) \rightarrow 0 \qquad (n,T \rightarrow \infty),
\end{equation*}
with  $\sigma^2 := \text{var}(v_{i,t})$  \citep[see][p. 1128, Lemma 2]{Alvarez2003}.  Using these results, they showed that if $\lim (T/n) = c$, with $ 0 \le c < \infty$, then
\begin{equation*}
	\sqrt{nT}\left( \widehat{\beta} - \beta\right) - \left[-  \sqrt{T/n}\left(1 + \beta\right)\right] \overset{d}{\rightarrow}  N(0,\, 1 - \beta^2) \qquad (n,T \rightarrow \infty). 
\end{equation*}
 \citep[see][p.1129, Theorem 2]{Alvarez2003}. Therefore, for the AR(1) panel data model, $\sqrt{nT}( \widehat{\beta} - \beta)$ has an asymptotic bias term of $\theta := - \sqrt{c}(1 + \beta)$, which is zero only when $c = 0$. 

\fussy
On the other hand, if $T/n \rightarrow c > 0$, then $\theta_{n,T} \overset{p}\rightarrow \theta \ne 0$, and 
\begin{equation} \label{ar1_normal_dist}
	\sqrt{nT}\left( \widehat{\beta} - \beta\right)  \overset{d}{\rightarrow}  N(\theta,\, 1 - \beta^2) \qquad (n,T \rightarrow \infty). 
\end{equation}
The fact that $\sqrt{nT}( \widehat{\beta} - \beta)$ has a bias term in its asymptotic distribution does not imply  $\widehat{\beta}$ is inconsistent. In fact, it is a consistent estimator of $\beta$ \citep[see][p.1129, Theorem 2]{Alvarez2003}.\footnote{This conclusion follows from (\ref{ar1_normal_dist}). This is because the limit distribution in (\ref{ar1_normal_dist}) implies that, for large $nT$, 
$\widehat{\beta}$ has an approximate normal distribution that is centered at $\beta +\theta/\sqrt{nT}$ with a variance of $(1-\beta)/(nT)$.  Hence, the limit distribution of $\widehat{\beta}$, as $n,T \rightarrow \infty$, is degenerate at $\beta$, which implies $\widehat{\beta}$ converges in probability to $\beta$.} However, because the asymptotic distribution of $\sqrt{nT}( \widehat{\beta} - \beta)$ is not centered at zero, large sample confidence intervals and test statistics will be misleading.  Specifically, the actual coverage of a confidence interval and the size of a test will differ from what they would be if the distribution of $\sqrt{nT}( \widehat{\beta} - \beta)$ were centered at zero.

The case considered by Alvarez and Arellano is instructive for two reasons. First, it illustrates that  $\theta_{n,T} \overset{p}\rightarrow 0$, as $n,T \rightarrow \infty$, is required for the asymptotic distribution of $\sqrt{nT}( \widehat{\beta} - \beta)$ to be centered at zero.  
The second reason the example is instructive is less obvious and is the point of the first result (Theorem \ref{bound_thm}) provided here. The  fact that $\theta_{n,T} \rightarrow \theta \ne 0$ when $c > 0$ is due to the number of instrumental variables that are used.  \cite{Alvarez2003} assumed all available moment restrictions are exploited, in which case  the maximum number of instrumental variables used in a period increases at the rate $T$ increases, and the total number of moment restrictions  increases at the rate $T^2$ increases.   It has long been known that using many moment restrictions has deleterious effects on the bias of a GMM estimator.  However, Theorem \ref{bound_thm}  sheds light on how many moment restrictions can be used without leading to large sample bias.  

\sloppy
Theorem \ref{bound_thm} provides conditions under which a generalization of $\theta_{n,T}$ converges in probability to a vector of zeros.  Specifically, let $\boldsymbol{A}_{n,T} :=  (1/(nT)) \sum_{t=1}^{T-1}\ddot{\boldsymbol{X}}_t'\boldsymbol{P}_t \ddot{\boldsymbol{X}}_t $ and $\boldsymbol{b}_{n,T}  :=  (1/\sqrt{nT}) \sum_{t=1}^{T-1}\ddot{\boldsymbol{X}}_t'\boldsymbol{P}_t \ddot{\boldsymbol{v}}_t$, with $ \ddot{\boldsymbol{v}}_{t} := c_{t} \left(   \boldsymbol{v}_{t} - \overline{\boldsymbol{v}}_{t} \right) $, $\boldsymbol{v}_t' := \left(v_{1,t}, \ldots, v_{n,t} \right) $, and $ \overline{\boldsymbol{v}}_t := \left( 1/(T-t)\right)\sum_{s=1}^{T-t} \boldsymbol{v}_{t+s} $. Also, set $\boldsymbol{\theta}_{n,T} := \boldsymbol{A}_{n,T}^{-1}E(\boldsymbol{b}_{n,T})$. Then, analogous to (\ref{bias_ar1}), we have
\begin{equation*} \label{bias_gen}
	\sqrt{nT}\left(\widehat{\boldsymbol{\beta}} - \boldsymbol{\beta} \right) -  \boldsymbol{\theta}_{n,T} =   \boldsymbol{A}_{n,T}^{-1}\left(\boldsymbol{b}_{n,T} - E(\boldsymbol{b}_{n,T})\right).
\end{equation*} 
For every $n$ and $T$, the distribution of $\boldsymbol{b}_{n,T} - E(\boldsymbol{b}_{n,T})$ is centered at a vector of zeros, $\boldsymbol{0}$.  It follows that if   $\boldsymbol{A}_{n,T} \overset{p}\rightarrow \boldsymbol{A} > 0$,\footnote{The notation $\boldsymbol{B} >0$, when $\boldsymbol{B} $ is a matrix, means $\boldsymbol{B} $ is a positive definite matrix.} the distribution of $\sqrt{nT}\left(\widehat{\boldsymbol{\beta}} - \boldsymbol{\beta} \right) -  \boldsymbol{\theta}_{n,T}$ becomes centered at $\boldsymbol{0}$ as the sample size grows. Therefore, whether or not the distribution of $\sqrt{nT}(\widehat{\boldsymbol{\beta}} - \boldsymbol{\beta} )$ is  centered at $\boldsymbol{0}$ depends on whether or not $\boldsymbol{\theta}_{n,T}  \overset{p}\rightarrow \boldsymbol{0}$.  If the distribution of  $\sqrt{nT}(\widehat{\boldsymbol{\beta}} - \boldsymbol{\beta} )$ is not centered at $\boldsymbol{0}$ in large samples, then the estimator $\widehat{\boldsymbol{\beta}}$ will henceforth be described as having asymptotic bias.  Theorem \ref{bound_thm} shows that whether or not an FOD GMM estimator has asymptotic bias  depends not just on the relative sizes of $n$ and $T$ but also on the number of instrumental variables used per period.  

\fussy
A few more definitions are needed in order to state Theorem \ref{bound_thm}.  Let $ \boldsymbol{w}_{i,t} $ be a column vector consisting of all of the distinct entries in $ \left\lbrace  \left( \boldsymbol{x}_{i,s}', \boldsymbol{z}_{i,s}'\right)  ; \; s = 1,\ldots, t  \right\rbrace $. That is, $ \boldsymbol{w}_{i,t} $ contains all of the distinct values of the explanatory variables and instrumental variables for the $i$th individual from the first period up to the $ t $th period.  Also, set $ \boldsymbol{u}_{i}' := (\boldsymbol{w}_{i,T}',v_{i,1},\ldots, v_{i,T},\eta_i ) $, and let $\gamma_{k,t,s} := \text{cov}\left(   v_{1,t} , x_{1,t+s,k} |\boldsymbol{w}_{1,t} \right)$.

In addition to these definitions, Theorem \ref{bound_thm} relies on several conditions:

	\begin{itemize}
			\item [A1:] the  $ \boldsymbol{u}_i $s are independent and identically distributed $($i.i.d.$)$ across $ i $;
			\item [A2:] $ \text{rank}\left(\boldsymbol{Z}_t\right) = q_t  $ with probability 1 (wp1);
		\item [A3:] $ E(v_{1,t}|\boldsymbol{w}_{1,t}) = 0 $; 
		\item[A4:] $\boldsymbol{A}_{n,T} \overset{p}\rightarrow \boldsymbol{A} > 0$; and
		\item [A5:] $ \left| \sum_{s=1}^{S} \gamma_{k,t,s} \right| \le M  $ wp1, for some finite $M $ and all $t  \ge 1$, $S \ge 1$, and $k=1,\ldots,K$.
	\end{itemize}

These conditions are weak enough to be satisfied by many dynamic regressions. For example, the estimation problems studied by \cite{Alvarez2003} and \cite{Hsiao2017}, among others \citep[e.g.,][]{Bun2006,Okui2009}, are special cases of the estimation problem examined here. 

The first four assumptions are relatively straightforward.  Nevertheless, it is worthwhile to review them to clarify what is---and what is not---required.  For example, Assumption A1 does not require that the errors be conditionally homoskedastic or time series homoskedastic.  Moreover, the i.i.d. assumption across $i$ can possibly be relaxed, but unfortunately relaxing the assumption would appear to require more technical detail. 

The second assumption implies $n$ must increase with $T$ if the largest number of instrumental variables per period increases with $T$. Specifically, it implies $n \ge q_T^{\ast}$, and, therefore, if $q_T^{\ast}$ increases with $T$,  then $n$ must likewise increase at least as fast as $q_T^{\ast}$ increases.  On the other hand, Assumption A2 does not rule out cases where $n$ is fixed.  For example, if at most $q$ instruments are used each period and $q$ does not increase with $T$, then $n$ can be fixed while $T$ increases provided $n \ge q$. 

The third assumption implies the error in period $t$ ($v_{i,t}$) is uncorrelated with the current and past explanatory variables.  The assumption also implies the current error is uncorrelated with the instruments used in period $t$. However, it imposes no additional restrictions on the instruments.  The entries in $\boldsymbol{z}_{i,t}$  can consist of, but need not be restricted to, current and past explanatory variables.  Moreover, they  can be in levels, differenced, or transformed in some other way.  Regardless of what instrumental variables are used or how they are constructed, Assumption A3 allows for dynamic panel data regressions with additional predetermined explanatory variables, provided the $v_{i,t}$s are uncorrelated across time.  

\sloppy
As for the fourth assumption, it implicitly entails some familiar conditions which are typically taken for granted.  To see this, first note that Assumption A2 ensures $\left[(1/n)\boldsymbol{Z}_t'\boldsymbol{Z}_t\right]^{-1}$ is defined (wp 1).  This, in turn, implies $\boldsymbol{\Psi}_{n,t} := (1/n)\ddot{\boldsymbol{X}}_t'\boldsymbol{Z}_t\left[(1/n)\boldsymbol{Z}_t'\boldsymbol{Z}_t\right]^{-1}(1/n)\boldsymbol{Z}_t'\ddot{\boldsymbol{X}}_t$ and  $\boldsymbol{A}_{n,T} = (1/T)\sum_{t=1}^{T-1}\boldsymbol{\Psi}_{n,t}$ are defined. But in order for $\boldsymbol{\Psi}_{n,t}$ to be positive definite,  the rank of  $(1/n)\boldsymbol{Z}_t'\ddot{\boldsymbol{X}}_t$ must be $K$.  That, in turn, requires the well-known necessary condition that $q_t \ge K$. In other words, it requires that we use at least as many instrumental variables each period as explanatory variables.  Moreover, if the $\boldsymbol{\Psi}_{n,t}$s are positive definite, then the average of these positive definite matrices---that is, $\boldsymbol{A}_{n,T} $---is positive definite.\footnote{Not all of the $\boldsymbol{\Psi}_{n,t}$s need be positive definite in order for $\boldsymbol{A}_{n,T} $ to be positive definite.}. All of this is typically taken for granted, because if $\boldsymbol{A}_{n,T} $ is not positive definite, then $\widehat{\boldsymbol{\beta}}$ is not defined.   Therefore, it is not unreasonable to assume the average of the $\boldsymbol{\Psi}_{n,t}$s ($\boldsymbol{A}_{n,T}$) converges (in probability), and that the limit ($\boldsymbol{A}$) is---like $\boldsymbol{A}_{n,T}$---positive definite.  An example in which the condition is satisfied is provided by  \cite{Alvarez2003}.  

\fussy
The fifth assumption is a characterization of weak dependence.  It is likely unfamiliar, but, like the other conditions, it does not appear to impose significant limitations on the types of panel data models to which the conclusion of Theorem 1 applies.  Assumption A5 characterizes the linear association between the  error in period $t$ and the $ k $th regressor $ s $ periods in the future relative to that period, conditional on the available information at time $t$.  Because the conditional covariance between $ v_{1,t} $ and $ x_{1,t+s,k} $ is not restricted to be zero, Assumption A5 allows for predetermined regressors.  Moreover, if the error term and future values of the explanatory variables are suitably weakly dependent, the bound on  the sum of covariances in A5 will be satisfied. 

This fact is illustrated by a stationary $K$th order autoregressive (AR($K$)) panel data model; see Lemma \ref{AR_K_lemma}.  Proofs are provided in the appendix.

\begin{lemma} \label{AR_K_lemma}
Let
\begin{equation*}
	\beta(L) y_{1,t} =  \eta_1 + v_{1,t} \qquad (t = 0, \pm 1,  \pm 2, \ldots),
\end{equation*}
where $ \beta(L) := 1 - \beta_1 L -  \beta_2 L^2 - \cdots - \beta_K L^K$, and $ L $ is the lag operator.  Also, let $ \boldsymbol{w}_{1,t}' = (y_{1,1-K},  \ldots,  y_{1,t-1}) $  $( t=1,2,\ldots,T  )$.
Assume the roots of the characteristic equation $ 1 - \beta_1 z - \beta_2 z^2 - \cdots - \beta_K z^K = 0 $ all lie outside the unit circle. Also, assume the    $v_{1,t}$s are i.i.d. across $ t $, with mean $ 0 $ and variance $ \sigma^2 $,   and $\eta_1 $ and $v_{1,t}$ are independent for all  $t$.   Then Assumption A5 is satisfied.\footnote{Because variables are assumed to be identically distributed across $i$, this lemma is stated in terms of a representative cross-sectional unit---the first ($i=1$).} 
\end{lemma}

The AR(K) panel data model is just one example of the models that satisfy Assumption A5.  In the Supplemental Appendix \citep{Phillips2024}, I show that A5 is satisfied by a variety of models.  These models include---but are not limited to---panel vector autoregressions and other panel data models with predetermined regressors.  The panel data models considered by \cite{Alvarez2003}, \cite{Bun2006}, \cite{Hsiao2017}, and \cite{Okui2009}, among others, are special cases of models that satisfy A5.

Theorem \ref{bound_thm} can now be stated. 

\begin{theorem} \label{bound_thm}  
	Assume A1 through A5 are satisfied and that $ T \ge 3 $. If  $ q_{T}^{\ast} = O(T^{\alpha}) $, 
then $\boldsymbol{\theta}_{n,T} \overset{p}\rightarrow \boldsymbol{0}$ as $T^{2\alpha}(\ln T)^2/(nT) \rightarrow 0$.
\end{theorem}

Theorem \ref{bound_thm} shows that the absence of asymptotic bias depends on the relative rate of increase in $n$ and $T$ and on the largest number of instrumental variables used per period ($q_T^{\ast}$).  The maximum per-period number of instruments may depend on $T$. A bound on how fast $q_T^{\ast}$ increases with  $T$ is parameterized in the theorem by $\alpha$.  

The case $\alpha = 1$ has been widely studied in the literature.  If all available  instrumental variables are used, and past lags of explanatory variables are viable instruments, then the maximum number of instruments used in a period increases with $T$ at the rate $T$ increases.  This corresponds to $\alpha = 1$. Then $T^{2\alpha}(\ln T)^2/(nT) = T(\ln T)^2/n$, which clearly does not go to zero for all sequences of  $n$ and $T$.  Instead, it only goes to zero for sequences of  $n$ and $T$ for which either $T$ does not increase or it increases slowly enough relative to $n$ that $T(\ln T)^2/n \rightarrow 0$.  The set of sequences for which $T(\ln T)^2/n \rightarrow 0$ is smaller than what \cite{Alvarez2003} found, for they found that there was no bias term if $T/n \rightarrow 0$.   However, \cite{Alvarez2003} established their result for an AR(1) panel data model with i.i.d. homoskedastic errors. Theorem \ref{bound_thm}, on the other hand, provides sufficient conditions for how fast $T$ can increase relative to $n$  for more general panel data regressions and under weaker assumptions about the errors.

Moreover, although $\boldsymbol{\theta}_{n,T} \overset{p}\rightarrow \boldsymbol{0}$ is only guaranteed for some sequences of $n$ and $T$ when $\alpha = 1$, we get $T^{2\alpha}(\ln T)^2/(nT) \rightarrow 0$ for all sequences of $n$ and $T$ if $\alpha < 1/2$.  Hence, Theorem \ref{bound_thm} shows that $\sqrt{nT}(\widehat{\boldsymbol{\beta}} - \boldsymbol{\beta})$ has no asymptotic bias regardless of what happens to $T/n$, provided the maximum number of instrumental variables used in a period increases with $T$ more slowly than $T^{1/2}$ increases.

An important special case that satisfies the restriction $\alpha < 1/2$  is when the number of instrumental variables used per period never exceeds some fixed number, say $q$.  Then, $\alpha = 0$.  Thus, if a researcher uses a fixed number of instruments each period, the FOD GMM estimator has no asymptotic bias regardless of how $n$ and $T$ increase. In fact,  although it is unusual for $n$ to be small while $T$ is large, Theorem \ref{bound_thm} nevertheless tells us there will be little bias in the distribution of $\sqrt{nT}(\widehat{\boldsymbol{\beta}} - \boldsymbol{\beta})$ even when $n$ is relatively small provided $T$ is large and $n \ge q$.   This conclusion is similar to what is true for the fixed effects estimator when the explanatory variables are all strictly exogenous.  That estimator, like the FOD GMM estimator, relies on eliminating fixed effects by subtracting from each variable a within cross-sectional average.     

\subsection{A sequential limit result}
The absence of bias in the distribution of $\sqrt{nT}(\widehat{\boldsymbol{\beta}} - \boldsymbol{\beta})$ tells us only that.  It does not tell us what distribution approximates the distribution of $\sqrt{nT}(\widehat{\boldsymbol{\beta}} - \boldsymbol{\beta})$ when $n$ and $T$ are both large.  To address that question, \cite{Hsiao2017} studied a choice of $q_t$ for which $\alpha = 0$ in some detail.  Specifically, they obtained asymptotic distribution results for FOD GMM estimators of the autoregression parameter in the AR(1) panel data model  using a single instrumental variable per period.  In doing so, they used sequential limits---specifically, they let $n \rightarrow \infty$, and then $T \rightarrow \infty$.\footnote{An alternative sequential-limit approach is to take limits in the order $T \rightarrow \infty$, then $n \rightarrow \infty$ (see, e.g., Phillips and Moon, 1999; Phillips and Moon, 2000; Kapetanios, 2008).} 

Phillips and Moon remark that  ``Sequential limit theory is easy to derive and generally leads to quick  results for a variety of model configurations'' \cite[][p. 1059]{Moon1999}.  Moreover, sequential limit results can sometimes be obtained under weaker assumptions than those needed for joint limit results \citep{Moon1999, Moon2000}.  These advantages  may explain why sequential limits have been used to analyze estimator properties when both $n$ and $T$ are large  \citep[see, e.g.,][]{Kapetanios2008, Hsiao2015, Hsiao2017}. 

However, taking  limits sequentially has a drawback: sequential limit results are not always robust.  Specifically, a sequential limit is not, in general, guaranteed to be equal to a joint limit \citep{Moon2000}.\footnote{\cite{Moon1999} provide conditions that, if satisfied, allows sequential limit results to be strengthened to joint limit results.  However, their paper focuses on taking sequential limits with $T \rightarrow \infty$, and then $n \rightarrow \infty$. Moreover, even for limits are taken in this order, it can be difficult to verify the conditions that imply a sequential limit equals a joint limit.}   Indeed, in the present application, letting $n \rightarrow \infty$ first, and then $T \rightarrow \infty$, always leads to the conclusion that there is no asymptotic bias in the distribution of $\sqrt{nT}(\widehat{\boldsymbol{\beta}} - \boldsymbol{\beta})$ regardless of $q_T^{\ast}$.  This is illustrated by Theorem \ref{normality_thm}

Before stating Theorem \ref{normality_thm}, some additional assumptions and definitions are needed.  Specifically,  assume
	\begin{itemize}
		\item [] A6: the entries in $\boldsymbol{x}_{1,t}$ have finite second-order moments, while $\ddot{v}_{1,t}$ and the entries in $ \boldsymbol{z}_{1,t} $  have finite fourth-order moments for all $t$.
	\end{itemize}
Assumption A6 implies the entries in the matrix $\boldsymbol{C}_t := E(\boldsymbol{z}_{1,t} \ddot{\boldsymbol{x}}_{1,t}^{\prime})$ are finite.  Similarly, the entries in the matrices $\boldsymbol{Q}_t := E(\boldsymbol{z}_{1,t} \boldsymbol{z}_{1,t}^{\prime})$, and $E(\ddot{v}_{1,s}\ddot{v}_{1,t}\boldsymbol{z}_{1,s}\boldsymbol{z}_{1,t}')$ are finite.  Moreover, given Assumption A2 implies $\boldsymbol{Q}_t$ is positive definite, if Assumptions A6 and A2 are both satisfied, we can define $\boldsymbol{\Sigma}_{t,t+s} := \boldsymbol{\Pi}_t^{\prime}E(\ddot{v}_{1,t}\ddot{v}_{1,t+s}\boldsymbol{z}_{1,t}\boldsymbol{z}_{1,t+s}')\boldsymbol{\Pi}_{t+s}$, where $\boldsymbol{\Pi}_t := \boldsymbol{Q}_t^{-1} \boldsymbol{C}_t$.   Also, let $\boldsymbol{\Omega}_{T} = (1/T)\sum_{t=1}^{T-1}\boldsymbol{\Sigma}_{t,t} + (1/T)\sum_{t=1}^{T-2}\sum_{s=1}^{T-1-t}\left( \boldsymbol{\Sigma}_{t,t+s} + \boldsymbol{\Sigma}_{t+s,t}\right)$. Then the last set of assumptions for Theorem \ref{normality_thm} are 
	\begin{itemize}
		\item [] A7: $ \boldsymbol{\Omega} := \lim_{T\rightarrow \infty} \boldsymbol{\Omega}_T  $ exists; \\
		and
		\item [] A8: $\boldsymbol{A} := \lim_{T\rightarrow \infty}(1/T)  \sum_{t=1}^{T-1} \boldsymbol{C}_t ' \boldsymbol{Q}_t^{-1} \boldsymbol{C}_t > 0$ exists. \\
\end{itemize}

\fussy
\begin{theorem} \label{normality_thm}
	Assume  A1  through A3 and A6 through A8 are satisfied. 	Then 
	\begin{equation*}
		\sqrt{nT}\left(\widehat{\boldsymbol{\beta}} - \boldsymbol{\beta}\right) \overset{d}{\rightarrow} N\left(\boldsymbol{0},\boldsymbol{A}^{-1} \boldsymbol{\Omega}\boldsymbol{A}^{-1}\right) \qquad ( n, T \rightarrow \infty)_{\text{seq}} . \\
	\end{equation*} 
\end{theorem} 

The notation ``$ ( n, T \rightarrow \infty)_{\text{seq}} $'' means $ n \rightarrow \infty $, then $ T \rightarrow \infty $. 

Theorem \ref{normality_thm}  illustrates the problem with taking limits sequentially in the present context.  Specifically, Theorem \ref{normality_thm} says the asymptotic distribution of $\sqrt{nT}(\widehat{\boldsymbol{\beta}} - \boldsymbol{\beta})$ is centered at $\boldsymbol{0}$, and it does so without referring to how many instrumental variables are used.  However, as has already been noted, \cite{Alvarez2003} showed that if all instrumental variables are used, the FOD GMM estimator they considered has a nonzero bias term in its asymptotic distribution if  $T/n \rightarrow c >0$, as $n, T \rightarrow \infty$. Alvarez and Arellano examined estimation of the AR(1) panel data model, which is a special case of the models covered by Theorem \ref{normality_thm}.  Therefore, Theorem \ref{normality_thm} contradicts the result provided in \cite{Alvarez2003}.

The contradiction is not the fault of Alvarez and Arellano. Instead, it arises from how limits are executed.  Taking a limit with $n \rightarrow \infty$ first,  implicitly holds $T$ fixed initially,  and GMM estimators do not generally have asymptotic bias, as $n \rightarrow \infty$, in the fixed $T$ case.  Hence, the bias term is gone by the time the second step---which consists of letting $T \rightarrow \infty$---is reached.

On the other hand, under the conditions of Theorem \ref{bound_thm}, there is also no bias term regardless of how $n$ and $T$ increase, provided the maximum number of per-period instruments, $q_T^{\ast}$, is selected so that $\alpha < 1/2$.  In the next section, the accuracy of the normal approximation, for large $n$ and $T$ and $\alpha = 0$, is investigated with Monte Carlo experiments.

\section{Simulations \label{MC}}

\cite{Phillips2020} provides Monte Carlo evidence that supports the claim that the distribution of $\sqrt{nT}(\widehat{\boldsymbol{\beta}} - \boldsymbol{\beta})$ is centered at $\boldsymbol{0}$ when, for example, $\alpha = 0$ and $T$ is not quite small compared to $n$.  That paper reports Monte Carlo evidence illustrating that when the per-period number of instrumental variables is fixed, FOD GMM generally has less bias and is more efficient than one-step first difference GMM  (FD GMM). In those experiments, the FOD GMM estimator was also less biased and usually more efficient than a two-step FD GMM in the presence of heteroskedastic errors, provided $T$ is not particularly small (e.g., $T=40$, $n=200$).\footnote{See \cite{Arellano1991} for a description of one-step and two-step FD GMM.}  

This section, on the other hand, reports on Monte Carlo experiments that were used to investigate the reliability of confidence intervals based on the FOD GMM estimator.  When $q_T^{\ast}$ was chosen so that $\alpha = 0$, the coverage of the confidence intervals based on the FOD GMM estimator turned out to be consistent with the conclusion of Theorem \ref{normality_thm}.  Specifically, for these experiments, the coverage of the FOD GMM confidence intervals, which relied on the normal approximation in Theorem \ref{normality_thm}, was accurate for large $T$.   In addition to confidence intervals, this section compares the estimator's  precision, as measured by root mean squared error (RMSE), to the precision of the efficient GMM estimator.

\subsection{Samples}

Monte Carlo samples were generated using a sampling scheme similar to one of the sampling schemes described  in \cite{Bun2006}.\footnote{\cite{Bun2006} considered two sampling schemes, and sampling schemes similar to both were initially considered.  However, the results were qualitatively similar across the two schemes.  Therefore, for the sake of brevity, the results for only one of them are reported here.}
In particular, the dependent variable in the regression model was generated according to
\begin{equation*}
y_{i,t}=\beta_1 y_{i,t-1}+\beta_2 x_{i,t}+\eta _{i}+v_{i,t}%
\qquad (t=-49,\ldots,-1,0,1,\ldots ,T,\ i=1,\ldots ,n),
\end{equation*}
with $ \beta_1 \in \left\lbrace 0.25,0.75\right\rbrace $ and $ \beta_2 = 1 - \beta_1 $.  The start-up value for  $ y_{i,-50} $ was  zero. Moreover, the error components $ v_{i,t} $ and $ \eta_i $  were generated independently of one another as i.i.d. standard normal variates. As for the explanatory variable $ x_{i,t} $, it  was generated as
\begin{equation*}
\begin{split}
x_{i,-50} & = \kappa_1\eta_i + \varepsilon _{i,-50},\\
x_{i,-49} & = \kappa_1\eta_i + \frac{1}{1 - \rho L}\left( \varepsilon_{i,-49} + \phi_1 v_{i,-50}\right) \\
x_{i,t} & = \kappa_1\eta_i + \frac{1}{1 - \rho L} \varepsilon_{i,t} + \phi_1 v_{i,t-1} \qquad (t = -48, \ldots, -1, 0, +1, \ldots, T),
\end{split}
\end{equation*}
with $ \rho \in \left\lbrace 0.50, 0.95 \right\rbrace  $, $ \kappa_1 \in \left\lbrace-1,0,+1 \right\rbrace  $, and $ \phi_1 \in \left\lbrace-1,0,+1 \right\rbrace  $.  Moreover, the $ \varepsilon_{i,t} $s were generated as $ \varepsilon_{i,t} \overset{i.i.d.}{\sim} U(-\sqrt{12}/2,\,\sqrt{12}/2)$ independently of the $ v_{i,t} $s and $ \eta_i $s.

I considered a total of 36 experimental designs, where a design is a combination of parameter values.  Table \ref{designs} lists Designs 1 through 18.  For these designs, I set $ \beta_1 = 0.25 $. Designs 19 through 36 were identical to Designs 1 through 18, except that  $ \beta_1 $ was set to $0.75$ for Designs 19 through 36.  The latter designs will henceforth be described as the weak instruments designs, for it is well-known that lagged values of the dependent variable become weaker instruments as $\beta_1$ approaches one  \citep[see, e.g.,][]{Blundell1998}.

\begin{table}[t]

	\caption{Designs 1 through 18 ($ \beta_1 = 0.25)$.}
\begin{center}
	\begin{tabular}{lrrrrrrrrr}
		\hline
		&  &  &  &  &  &  &  &  &  \\
		
		&  \multicolumn{9}{c}{Designs}  \\
		& 1 & 2 & 3 & 4 & 5 & 6 & 7 & 8 & 9 \\
		&  &  &  &  &  &  &  &  &  \\
		
		$ \rho $   & $ 0.50 $ & $ 0.50 $ & $ 0.50 $ & $ 0.50 $ & $ 0.50 $ & $ 0.50 $ & $ 0.50 $ & $ 0.50 $ & $ 0.50 $ \\
		$ \phi_1 $	& $ -1.00 $ & $ -1.00 $ & $ -1.00 $ & $ 0.00 $ & $ 0.00 $ & $ 0.00 $ & $ 1.00 $ & $ 1.00 $ & $ 1.00 $  \\
		$ \kappa_1 $ & $ -1.00 $ & $ 0.00 $ & $ 1.00 $ & $ -1.00 $ & $ 0.00 $ & $ 1.00 $ & $ -1.00 $ & $ 0.00 $ & $ 1.00 $ \\
		&  &  &  &  &  &  &  &  &  \\

		&  \multicolumn{9}{c}{Designs}  \\
		& 10 & 11 & 12 & 13 & 14 & 15 & 16 & 17 & 18 \\
		&  &  &  &  &  &  &  &  &  \\
		
		$ \rho $   & $ 0.95 $ & $ 0.95 $ & $ 0.95 $ & $ 0.95 $ & $ 0.95 $ & $ 0.95 $ & $ 0.95 $ & $ 0.95 $ & $ 0.95 $ \\
		$ \phi_1 $	& $ -1.00 $ & $ -1.00 $ & $ -1.00 $ & $ 0.00 $ & $ 0.00 $ & $ 0.00 $ & $ 1.00 $ & $ 1.00 $ & $ 1.00 $  \\
		$ \kappa_1 $ & $ -1.00 $ & $ 0.00 $ & $ 1.00 $ & $ -1.00 $ & $ 0.00 $ & $ 1.00 $ & $ -1.00 $ & $ 0.00 $ & $ 1.00 $ \\

		&  &  &  &  &  &  &  &  &  \\
		\hline
		
	\end{tabular}
\end{center}
	\label{designs}
	
\end{table}

\sloppy
The effect of how each series was initialized was eliminated by dropping the first 50 values of each generated time series. As a result, estimation was based  on the values $(x_{i,0},y_{i,0}),(x_{i,1},y_{i,1}),\ldots, (x_{i,T},y_{i,T}) $  ($ i = 1,\ldots,n $).  Moreover,  I set $n = 200$. The number of time periods, $T$,  was set equal to 20, then 40, and finally 100 in order to examine the effect of increasing $T/n$ on confidence intervals and estimator precision.  Finally, for each  sample size and parameter design, 5,000 samples were generated. Estimates were calculated for each of these 5,000 samples.\footnote{The data were generated using GAUSS.  Moreover, all calculations were performed with GAUSS.}

\subsection{Estimators}

 After a sample was generated, three estimates of $\boldsymbol{\beta} = (\beta_1, \beta_2)'$ were calculated: an FOD GMM estimate, an FD GMM estimate, and an efficient GMM estimate.  For the FD and FOD GMM estimates,  I set $ \boldsymbol{z}_{i,1}^{\prime} = (y_{i,0},x_{i,0},x_{i,1}) $ and $ \boldsymbol{z}_{i,t}^{\prime} = (y_{i,t-2},y_{i,t-1}, x_{i,t-2},x_{i,t-1},x_{i,t}) $ ($ t=2,\ldots,T-1 $, $ i=1,\ldots,n $).  
Given these instrumental variables, FOD GMM estimates were calculated using the formula in (\ref{fod_est}).

The FD GMM estimates, on the other hand, were calculated using the formula
\begin{equation} \label{fd_est}
\begin{split}
\widetilde{\boldsymbol{\beta}}  := & \left[\sum_{i=1}^n\tilde{\boldsymbol{X}}_i'\boldsymbol{Z}_{d,i}\left( \sum_{i=1}^n\boldsymbol{Z}_{d,i}'\boldsymbol{G}\boldsymbol{Z}_{d,i}\right)^{-1} \sum_{i=1}^n \boldsymbol{Z}_{d,i}'\tilde{\boldsymbol{X}}_i  \right]^{-1} \\
& \times \sum_{i=1}^n \tilde{\boldsymbol{X}}_i'\boldsymbol{Z}_{d,i}\left( \sum_{i=1}^n\boldsymbol{Z}_{d,i}'\boldsymbol{G}\boldsymbol{Z}_{d,i}\right)^{-1} \sum_{i=1}^n \boldsymbol{Z}_{d,i}'\tilde{\boldsymbol{y}}_i.
\end{split}
\end{equation}
In this formula, $ \tilde{\boldsymbol{y}}_i' := (y_{i,2} - y_{i,1},\ldots,y_{i,T} - y_{i,T-1}) $;  $ \tilde{\boldsymbol{X}}_i $ stacks $ \tilde{\boldsymbol{x}}_{i,t+1}' := \boldsymbol{x}_{i,t+1}' - \boldsymbol{x}_{i,t}' $, with $\boldsymbol{x}_{i,t}' =  (y_{i,t-1},x_{i,t})$ ($ t=1,\ldots,T-1 $); and $\boldsymbol{Z}_{d,i}$ is defined in (\ref{Z_di}). Also, $ \boldsymbol{G} $ is a $ (T-1) \times (T-1) $ matrix with twos running down the main diagonal, minus ones just above and below the main diagonal, and zeros everywhere else \citep[see, e.g.,][]{Arellano1991}.  

Finally, efficient GMM estimates were also calculated. The efficient GMM estimator exploited all available instrumental variables.  For this estimator, I set $ \boldsymbol{z}_{i,1}^{\prime} = (y_{i,0},x_{i,0},x_{i,1}) $ and $ \boldsymbol{z}_{i,t}^{\prime} = (\boldsymbol{z}_{i,t-1}',y_{i,t-1},x_{i,t}) $ ($ t=2,\ldots,T-1 $, $ i=1,\ldots,n $).  To calculate efficient estimates, either the formula in (\ref{fod_est}) or the formula in (\ref{fd_est}) can be used, because when all available instrumental variables are used, the two formulas give numerically identical results  \citep[see, e.g.,][]{Phillips2019}.    Although (\ref{fod_est}) and (\ref{fd_est}) give the same estimate in this case, the formula in (\ref{fod_est}) was used to calculate estimates because it is more efficient computationally \citep{Phillips2020}. Moreover, efficient estimates were not calculated for $T=100$, because the requirement that $n$ be no smaller than $q_t$ was not satisfied---and hence $(\boldsymbol{Z}_t'\boldsymbol{Z}_t)^{-1}$ was not defined---for all $t$ when $T$ is this large relative to $n$ and all available instruments are used.  Finally, because FD GMM and FOD GMM are the same when all available instruments are used, the efficient GMM estimator will henceforth be referred to as the FD/FOD GMM estimator.

\fussy
In order to construct confidence intervals, standard errors were also calculated.  Their calculation was simplified by the fact that the experimental $v_{i,t}$s were conditionally homoskedastic and uncorrelated.  In this case, the variance-covariance matrix of $\sqrt{nT} (\widehat{\boldsymbol{\beta}}-\boldsymbol{\beta})$  simplifies to $\sigma^2 \boldsymbol{A}^{-1}$.  Therefore, standard errors for the  FOD GMM estimates were calculated by taking the square roots of the diagonal entries of $ \widehat{\sigma}^2\left( \sum_{t=1}^{T-1} \ddot{\boldsymbol{X}}_t'\boldsymbol{P}_t \ddot{\boldsymbol{X}}_t \right) ^{-1} $, where $ \widehat{\sigma}^2 :=\sum_{t=1}^{T-1}(\ddot{\boldsymbol{y}}_{t} - \ddot{\boldsymbol{X}}_{t}\widehat{\boldsymbol{\beta}})'(\ddot{\boldsymbol{y}}_{t} - \ddot{\boldsymbol{X}}_{t}\widehat{\boldsymbol{\beta}})/[n(T-1)] $.  Moreover, when all available instruments were used, standard errors for FD/FOD estimates were calculated similarly.  On the other hand, the  standard errors for the FD estimator, $ \widetilde{\boldsymbol{\beta}} $, were calculated using the square roots of the diagonal entries in $ \widetilde{\sigma}^2 \left[\sum_{i=1}^n \tilde{\boldsymbol{X}}_i'\boldsymbol{Z}_{d,i}\left( \sum_{i=1}^n\boldsymbol{Z}_{d,i}'\boldsymbol{G}\boldsymbol{Z}_{d,i}\right)^{-1} \sum_{i=1}^n \boldsymbol{Z}_{d,i}'\tilde{\boldsymbol{X}}_i  \right]^{-1} $, 
where $ \widetilde{\sigma}^2 := \sum_{i=1}^n( \tilde{\boldsymbol{y}}_{i}  -  \tilde{\boldsymbol{X}}_{i}\widetilde{\boldsymbol{\beta}})'( \tilde{\boldsymbol{y}}_{i}  -  \tilde{\boldsymbol{X}}_{i}\widetilde{\boldsymbol{\beta}}) /[2n(T-1)]  $.

\subsection{Simulation results}

\subsubsection{Confidence intervals}

Tables \ref{ci_beta1_1to18} through \ref{ci_beta2_19to36} report the coverage of 95 percent confidence intervals for $\beta_1$ and $\beta_2$ using the FOD, FD, and FD/FOD GMM estimators, their respective standard errors, and the normal approximation---specifically, each interval extends from 1.96 standard errors below a regression parameter estimate to 1.96 standard errors above it. Tables \ref{ci_beta1_1to18} and \ref{ci_beta2_1to18} provide the coverage of the confidence intervals for $\beta_1$ and $\beta_2$, respectively, for Designs 1 through 18, and Tables \ref{ci_beta1_19to36} and \ref{ci_beta2_19to36} provide the coverage of the confidence intervals for $\beta_1$ and $\beta_2$ for Designs 19 through 36.

\begin{table}[!]
	\caption{Coverage of 95-percent confidence intervals for $ \beta_1$, Designs 1--18 ($n=200$).}
	
	\begin{tabular}{lrcccccccccc}
		\hline
		&  &  &  & &  &  &  &  &  & & \\
		
	Estimator &  $T$  &  & & & & & & & &  & \\
		
		&  & Designs:    & 1 & 2 & 3 & 4 & 5 & 6 & 7 & 8 & 9 \\
		
		 &  &  &  &  &  &  &  &  & & & \\	
		 FD/FOD & 20 &  & 95.0 & 95.7 & 94.5 & 90.6 & 90.9 & 91.5 & 90.3 & 89.0 & 88.1 \\	
		 &  &  &  &  &  &  &  &  & & & \\
		FD:  & $ 20  $ & & 95.2 & 95.3 & 94.5 & 93.3 & 94.1 & 94.0 & 93.8 & 93.3 & 92.9 \\
		FD:  & $ 40 $ & & 94.7 & 95.0 & 94.0 & 93.4 & 93.8 & 92.8 & 93.0 & 92.7 & 92.4 \\
		FD &  $ 100$  & & 95.2 & 94.9 & 94.7 & 91.1 & 90.2 & 91.5 & 89.8 & 89.0 & 88.8\\		
		   &  &  &  &  &  &  &  &  & & & \\			
		FOD: &  $ 20  $& & 95.0 & 95.4 & 94.7 & 94.7 & 95.4 & 94.6 & 95.2 & 94.5 & 94.5 \\
		FOD: & $ 40 $ & & 94.6 & 95.0 & 94.8 & 94.9 & 95.6 & 94.8 & 94.8 & 95.1 & 94.5 \\
		FOD &  $ 100$  & & 95.1 & 94.9 & 94.9 & 94.5 & 94.8 & 95.2 & 94.9 & 94.9 & 95.0 \\
		&  &  &  &  &  &  &  &  & & & \\
		& & Designs:    & 10 & 11 & 12 & 13 & 14 & 15 & 16 & 17 & 18 \\
		 & &  &  &  &  &  &  &  &  &  & \\
		 FD/FOD& 20 &  & 95.0 & 95.4 & 94.7 & 91.9 & 91.2 & 91.0 & 90.7 & 90.4 & 89.7 \\
		 &  &  &  &  &  &  &  &  & & & \\
		 FD: & $ 20  $ & & 94.9 & 94.6 & 94.9 & 94.2 & 94.5 & 94.1 & 94.3 & 93.2 & 93.9 \\
		FD: & $ 40 $ & & 94.4 & 94.5 & 95.4 & 93.1 & 93.6 & 94.2 & 93.4 & 92.1 & 93.1 \\
		FD &  $ 100$  &  & 94.3 & 95.2 & 95.0 & 92.9 & 92.5 & 92.7 & 90.9 & 91.1 & 91.0 \\
		&  &  &  &  &  &  &  &  & & \\
		FOD:  & $ 20  $ & & 95.2 & 95.5 & 94.7 & 95.1 & 95.3 & 94.6 & 94.6 & 94.3 & 94.9 \\
		FOD:& $ 40 $ & & 94.9 & 95.3 & 95.4 & 95.0 & 94.8 & 95.4 & 95.1 & 94.7 & 94.4 \\
		FOD &  $ 100$  &  & 94.7 & 95.2 & 95.4 & 94.9 & 94.7 & 94.7 & 94.9 & 94.4 & 94.6 \\
		&  &  &  &  &  &  &  &  & & & \\
				
		\hline
	\label{ci_beta1_1to18}		
	\end{tabular} 
	
	Note: Each coverage estimate is based on 5,000 samples.
	
\end{table}

\begin{table}[!]
	\caption{Coverage of 95-percent confidence intervals for $ \beta_2$, Designs 1--18 ($n=200$).}
	
	\begin{tabular}{lrcccccccccc}
		\hline
		&  &  &  & &  &  &  &  &  & & \\
		
	Estimator &  $T$  &  & & & & & & & &  & \\
		
		&  & Designs:    & 1 & 2 & 3 & 4 & 5 & 6 & 7 & 8 & 9 \\
		
		 &  &  &  &  &  &  &  &  & & & \\	
		 FD/FOD & 20 &  & 94.1 & 93.9 & 93.2 & 94.8 & 95.2 & 95.2 & 95.2 & 94.7 & 94.6 \\
		 &  &  &  &  &  &  &  &  & & & \\
		FD:  & $ 20  $ & & 94.8 & 94.5 & 94.7 & 95.3 & 94.7 & 94.9 & 95.4 & 94.7 & 94.6 \\
		FD:  & $ 40 $ & & 94.0 & 94.2 & 94.0 & 94.9 & 95.2 & 94.6 & 94.8 & 95.2 & 94.9 \\
		FD &  $ 100$  & & 93.1 & 93.3 & 93.4 & 94.5 & 94.9 & 94.5 & 95.0 & 95.3 & 95.6 \\			
		 &  &  &  &  &  &  &  &  & & & \\
		 FOD: &  $ 20  $ & & 95.1 & 94.8 & 94.8 & 95.7 & 95.3 & 94.8 & 95.4 & 94.9 & 94.5 \\
		FOD: & $ 40 $ & & 94.9 & 94.7 & 95.1 & 95.3 & 95.0 & 95.1 & 94.7 & 95.4 & 95.2 \\
		FOD &  $ 100$  & & 95.0 & 94.9 & 95.6 & 94.9 & 94.2 & 94.9 & 95.1 & 95.7 & 95.5 \\
		
		&  &  &  &  &  &  &  &  & & & \\
		& & Designs:    & 10 & 11 & 12 & 13 & 14 & 15 & 16 & 17 & 18 \\
		 & &  &  &  &  &  &  &  &  &  & \\
		 FD/FOD& 20 &  & 93.4 & 92.4 & 92.8 & 94.4 & 94.3 & 94.5 & 94.6 & 94.8 & 94.7 \\
		 &  &  &  &  &  &  &  &  & & & \\
		FD: & $ 20  $ & & 95.0 & 94.0 & 94.3 & 95.1 & 95.1 & 94.8 & 94.7 & 95.0 & 95.0 \\
		FD: & $ 40 $ & & 94.0 & 94.0 & 94.6 & 94.8 & 94.4 & 95.5 & 95.0 & 94.6 & 94.7 \\
		FD &  $ 100$  & & 93.8 & 93.7 & 93.6 & 94.5 & 94.5 & 94.6 & 94.6 & 94.9 & 94.4 \\
		 &  &  &  &  &  &  &  &  & & & \\
		 FOD:  & $ 20  $ & & 94.8 & 94.7 & 94.2 & 95.0 & 94.9 & 95.1 & 94.4 & 95.0 & 95.0 \\
		FOD:& $ 40 $ & & 95.0 & 95.5 & 94.4 & 94.7 & 95.0 & 95.5 & 95.3 & 94.7 & 94.9 \\
		FOD &  $ 100$  & & 95.3 & 95.1 & 95.0 & 95.1 & 94.8 & 94.8 & 95.0 & 94.9 & 94.1 \\
		
		&  &  &  &  &  &  &  &  & & & \\
				
		\hline
	\label{ci_beta2_1to18}		
	\end{tabular} 
	
	Note: Each coverage estimate is based on 5,000 samples.
	
\end{table}

\begin{table}[!]
	\caption{Coverage of 95-percent confidence intervals for $ \beta_1$, Designs 19--36 ($n=200$).}
	
	\begin{tabular}{lrcccccccccc}
		\hline
		&  &  &  & &  &  &  &  &  & & \\
		
	Estimator &  $T$  &  & & & & & & & &  & \\
				& & Designs:    & 19 & 20 & 21 & 22 & 23 & 24 & 25 & 26 & 27 \\
			&  &  &  &  &  &  &  &  & & & \\
		FD/FOD& 20 &  & 67.6 & 62.6 & 61.4 & 62.7 & 56.2 & 52.0 & 66.1 & 58.8 & 51.8 \\
		 &  &  &  &  &  &  &  &  & & & \\

		FD: & $ 20  $ & & 87.8 & 82.9 & 79.4 &  85.9 & 84.4 & 82.1 & 86.7 & 84.8 & 82.0 \\
		FD: & $ 40 $ & & 87.5 & 86.4 & 84.6 & 86.1 & 85.3 & 83.7 & 87.7 & 86.5 & 85.4 \\
		FD &  $ 100 $  &  & 83.1 & 83.1 & 82.9 & 80.7 & 79.2 & 79.7 & 81.8 & 81.2 & 81.3\\
	
		 &  &  &  &  &  &  &  &  & & & \\	
		FOD: & $ 20  $ & & 93.5 & 92.1 & 91.2 & 93.3 & 92.3 & 91.3 & 93.8 & 93.0 & 91.5 \\
		FOD: & $ 40 $ & & 94.3 & 94.3 & 94.0 & 94.0 & 93.8 & 93.7 & 94.6 & 93.9 & 93.7 \\
		FOD &  $ 100$  &  & 95.2 & 94.5 & 94.3 & 94.9 & 94.9 & 94.4 & 94.7 & 94.7 & 95.1 \\
		&  &  &  &  &  &  &  &  & & &  \\
		& & Designs:    & 28 & 29 & 30 & 31 & 32 & 33 & 34 & 35 & 36 \\
		  & &  &  &  &  & &  &  &  &  &  \\
		  FD/FOD& 20 &  & 82.7 & 81.8 & 81.7 & 70.8 & 72.2 & 70.6 & 69.1 & 67.5 & 65.3 \\
		 &  &  &  &  &  &  &  &  & & & \\
		FD: & $ 20  $ & & 92.4 & 91.5 & 91.6 & 89.6 & 89.9 & 89.2 & 89.7 & 89.0 & 88.4 \\
		FD:   & $ 40 $ & & 93.1 & 92.0 & 92.3 & 90.5 & 90.3 & 90.6 & 90.1 & 89.9 & 90.0 \\
		FD &  $ 100$  &  & 91.7 & 91.7 & 92.3 & 88.4 & 88.1 & 88.1 & 87.7 & 87.0 & 87.4 \\
		 & &  &  &  &  & &  &  &  &  &  \\
		FOD: & $ 20  $ & & 94.5 & 94.3 & 94.1 & 93.8 & 94.3 & 93.5 & 94.1 & 93.6 & 93.5 \\
		FOD: & $ 40 $ & & 94.6 & 94.9 & 94.6 & 94.0 & 94.0 & 94.8 & 94.6 & 94.2 & 94.5 \\
		FOD &  $ 100$  &  & 94.7 & 95.0 & 94.7 & 94.9 & 94.7 & 94.4 & 94.9 & 94.9 & 94.7 \\
		&  &  &  &  &  &  &  &  & & & \\		
		\hline
	\label{ci_beta1_19to36}		
	\end{tabular} 
	
	Note: Each coverage estimate is based on 5,000 samples.
	
\end{table}

\begin{table}[!]
	\caption{Coverage of 95-percent confidence intervals for $ \beta_2$, Designs 19--36 ($n=200$).}
	
	\begin{tabular}{lrcccccccccc}
		\hline
		&  &  &  & &  &  &  &  &  & & \\
		
	Estimator &  $T$  &  & & & & & & & &  & \\
				& & Designs:    & 19 & 20 & 21 & 22 & 23 & 24 & 25 & 26 & 27 \\
			&  &  &  &  &  &  &  &  & & & \\
		FD/FOD& 20 &  & 92.2 & 87.3 & 83.8 & 94.1 & 94.4 & 92.1 & 94.9 & 94.3 & 94.2 \\
		&  &  &  &  &  &  &  &  & & & \\
		FD: & $ 20  $ & & 92.2 & 88.0 & 83.6 & 94.9 & 93.5 & 91.7 & 95.4 & 95.0 & 94.0 \\
		FD: & $ 40 $ & & 93.6 & 91.6 & 90.5 & 95.0 & 94.6 & 93.9 & 94.5 & 94.9 & 95.1 \\
		FD &  $ 100 $ & & 92.4 & 92.3 & 91.6 & 95.2 & 95.0 & 95.0 & 94.9 & 94.5 & 94.4 \\

		 &  &  &  &  &  &  &  &  & & & \\
		FOD: & $ 20  $   &  & 94.8 & 93.0 & 92.3 & 94.9 & 94.6 & 94.2 & 95.3 & 94.9 & 94.7 \\
		FOD: & $ 40 $ & & 94.7 & 94.5 & 94.3 & 94.5 & 94.7 & 95.1 & 94.5 & 95.0 & 94.8 \\
		FOD &  $ 100$  & & 95.0 & 94.7 & 94.5 & 95.5 & 95.3 & 94.6 & 95.7 & 94.9 & 94.9 \\

		&  &  &  &  &  &  &  &  & & &  \\
		& & Designs:    & 28 & 29 & 30 & 31 & 32 & 33 & 34 & 35 & 36 \\
		  & &  &  &  &  & &  &  &  &  &  \\
		  FD/FOD& 20 &  & 94.2 & 94.7 & 94.7 & 93.9 & 93.9 & 93.5 & 94.4 & 94.7 & 94.6 \\
		  &  &  &  &  &  &  &  &  & & & \\
		FD: & $ 20 $ & & 94.6 & 93.9 & 94.4 & 94.6 & 94.1 & 93.9 & 94.4 & 94.9 & 94.5 \\
		FD:   & $ 40 $ & & 95.4 & 94.8 & 94.8 & 94.1 & 95.0 & 95.1 & 94.6 & 94.1 & 94.2 \\
		FD &  $ 100$  & & 94.7 & 94.7 & 95.0 & 94.2 & 94.8 & 94.8 & 93.8 & 94.2 & 94.8 \\

		 &  &  &  &  &  &  &  &  & & & \\
		 FOD: & $ 20  $ & & 94.8 & 94.9 & 95.3 & 95.2 & 94.9 & 94.4 & 94.5 & 95.0 & 94.9 \\
		FOD: & $ 40 $ & & 94.7 & 95.1 & 94.7 & 94.7 & 94.3 & 95.2 & 94.7 & 94.7 & 94.4 \\
		FOD &  $ 100$  & & 94.7 & 94.9 & 94.9 & 95.2 & 94.8 & 94.4 & 94.5 & 94.9 & 95.5 \\
		
		&  &  &  &  &  &  &  &  & & & \\		
		\hline
	\label{ci_beta2_19to36}		
	\end{tabular} 
	
	Note: Each coverage estimate is based on 5,000 samples.
	
\end{table}

Coverage estimates for the FD/FOD confidence intervals are only provided for $T=20$.  This is the best case for the FD/FOD GMM estimator because the estimator was developed for situations where  $T$ is small compared to $n$.  Nevertheless, even in this case the coverage of the FD/FOD confidence intervals is often unreliable.  For example, consider the FD/FOD GMM intervals for $\beta_1$ given weak instruments (Table \ref{ci_beta1_19to36}).  For Designs 19 through 36, the coverage of the FD/FOD GMM intervals for $\beta_1$ ranged from a low of 51.8 percent (Design 27) to a high of 82.7 percent (Design 28).  Unsurprisingly, increasing $T$ to 40 made for even worse coverage.  This result is consistent with Alvarez and Arellano's finding of non-zero asymptotic bias in the distribution of the FOD GMM estimator of the AR(1) panel data model when $T$ is not small compared to $n$ and all available moment restrictions are exploited.

Using fewer than all available instrumental variables improved the reliability of the confidence intervals, as expected, but by how much depended on how the data were transformed to remove fixed effects.  Usually, the coverage of the FOD confidence intervals approximated 95 percent at least as well, if not better, than did the FD confidence intervals.  Moreover, confidence intervals based on the FOD GMM estimator  were no less reliable for $T=100$ than for $T=20$.  Indeed, the reliability of the FOD confidence intervals appears to improve as $T$  increases relative to $n$; see, for example, the FOD confidence intervals for $\beta_1$ and Designs 19 through 36 (Table \ref{ci_beta1_19to36}).   On the other hand, the FD confidence intervals often under estimated 95 percent, and their reliability did not improve as $T$ was increased relative to $n$; see, for example, Table \ref{ci_beta1_19to36}.\footnote{The finding that the coverage of the FD confidence intervals did not improve as $T$ was increased is consistent with analytical results \cite{Hsiao2017} provide for FD GMM estimation of the AR(1) panel data model.}   

Moreover, using critical values from the standard normal distribution, 90 percent and 50 percent confidence intervals were also calculated.  For the sake of brevity, the coverage of these intervals is omitted and is only summarized here:  their reliability  was similar to that of the 95 percent confidence intervals.  For example, for large $n$ and $T$, the coverage of the  90 and 50 percent FOD confidence intervals  approximated 90  and 50 percent, respectively.  Apparently for these experiments, the FOD GMM estimator has little bias and the normal approximation works well when $n$ and $T$ are both large and $q=0$.  On the other hand, the  coverage of the 90 and 50 percent FD confidence intervals often under estimated 90 and 50 percent, especially for the weak instrument designs (Designs 19 through 36).

\subsubsection{Relative precision}

Although confidence intervals based on the FOD GMM estimator have superior coverage accuracy, the estimator sacrifices estimation efficiency to achieve this superior accuracy.  In the experiments considered here, the FD and FOD GMM estimators use at most five instrumental variables each period.  The FD/FOD GMM estimator, on the other hand, exploits all available moment restrictions.  Therefore, the FD and FOD GMM estimators are inefficient relative to the FD/FOD GMM estimator.  However, for $\alpha < 1/2$, the FOD GMM estimator does not have asymptotic bias, and, therefore, it can improve on the FD and FD/FOD GMM estimators in terms of precision, as measured by RMSE.  The Monte Carlo experiments were also used to investigate this possibility.

\pgfplotsset{
    myplotstyle/.style={
    legend style={draw=none, font=\small},
    legend cell align=left,
    legend pos=north east,
    ylabel style={align=center},
    xlabel style={align=center},
    scaled ticks=false,
    every axis plot/.append style={thick},
    },
}

\begin{figure}[!] 
 \begin{center}
 \caption{RMSE of Estimator of $\beta_1$ Over RMSE\\
 of FD/FOD Estimator  ($T=20$).}
 \begin{tikzpicture}
\begin{axis}[myplotstyle,
xlabel = {Design No. \\
\\},
ylabel = {Relative Precision},
legend pos = south west,
legend entries = {FD, FOD,  {}},
]
\addplot[red] coordinates {
(1, 1.3162483)
(2, 1.3291872)
(3, 1.3154811)
(4, 1.1142385)
(5, 1.0781134)
(6, 1.0530521)
(7, 1.0760446)
(8, 1.0395619)
(9, 0.99827955)
(10, 1.4414003)
(11, 1.4600044)
(12, 1.4658971)
(13, 1.0766580)
(14, 1.0569152)
(15, 1.0709716)
(16, 1.0521168)
(17, 1.0485550)
(18, 1.0095184)
(19, 1.1388053)
(20, 1.2736915)
(21, 1.5434043)
(22, 1.0531652)
(23, 0.98227961)
(24, 1.0309466)
(25, 1.1243848)
(26, 1.0273801)
(27, 0.96540231)
(28, 1.4209466)
(29, 1.4305981)
(30, 1.4503935)
(31, 1.0406304)
(32, 1.0506364)
(33, 1.0467636)
(34, 0.96615201)
(35, 0.95718947)
(36, 0.93375449)
};
\addplot[blue] coordinates {
(1, 1.0138454)
(2, 1.0646101)
(3, 1.1208158)
(4, 0.89433425)
(5, 0.94935287)
(6, 1.0433042)
(7, 0.86414674)
(8, 0.88884358)
(9, 0.92859536)
(10, 1.0404080)
(11, 1.0350441)
(12, 1.0547984)
(13, 0.94423097)
(14, 0.94741261)
(15, 1.0709716)
(16, 0.92806093)
(17, 0.93631289)
(18, 0.92766894)
(19, 0.80231560)
(20, 0.98375503)
(21, 1.1561082)
(22,  0.70275610)
(23, 0.74037560)
(24, 0.81104552)
(25, 0.67132053)
(26, 0.68500958)
(27, 0.71354337)
(28, 0.87047558)
(29, 0.87765614)
(30, 0.89956432)
(31, 0.76359997)
(32, 0.76636719)
(33, 0.77320690)
(34, 0.74504141)
(35, 0.75081340)
(36, 0.75608460)
 };
  \addplot[black] coordinates {
 (1, 1)
 (2,1)
 (3, 1)
 (4, 1)
 (5, 1)
 (6, 1)
 (7,1)
 (8, 1)
 (9, 1)
 (10, 1)
 (11, 1)
 (12, 1)
 (13, 1)
 (14, 1)
 (15,1)
 (16, 1)
 (17, 1)
 (18, 1)
 (19, 1)
 (20,1)
 (21, 1)
 (22, 1)
 (23, 1)
 (24, 1)
 (25, 1)
 (26, 1)
 (27, 1)
 (28, 1)
 (29, 1)
 (30,1)
 (31, 1)
 (32, 1)
 (33, 1)
 (34, 1)
 (35, 1)
 (36, 1)
  };
  \end{axis}
  \label{precision_beta1_T20}
\end{tikzpicture} 
\end{center}
 \end{figure}

\begin{figure}[!]
 \begin{center}
 \caption{RMSE of Estimator of $\beta_1$ Over RMSE\\ 
 of FD/FOD Estimator ($T=40$).}
 \begin{tikzpicture}
\begin{axis}[myplotstyle,
xlabel = {Design No. \\
\\},
ylabel = {Relative Precision},
legend pos = south west,
legend entries = {FD, FOD,  {}},
]
\addplot[red] coordinates {
(1, 1.4145625)
(2, 1.3795004)
(3, 1.4098140)
(4, 1.0605226)
(5, 1.0292027)
(6, 1.0157669)
(7, 0.99831579)
(8, 0.95885935)
(9, 0.92105941)
(10, 1.5972292)
(11, 1.5947444)
(12, 1.5690847)
(13, 1.0340525)
(14, 1.0298328)
(15, 1.0232780)
(16, 1.0070015)
(17, 0.99494932)
(18, 0.98640476)
(19, 1.1005127)
(20, 1.1052375)
(21, 1.2552138)
(22, 0.93961877)
(23, 0.87945147)
(24, 0.89514494)
(25, 0.91445016)
(26, 0.85314895)
(27, 0.78992992)
(28, 1.3695411)
(29, 1.4222590)
(30, 1.4413082)
(31, 0.93146194)
(32, 0.92922122)
(33, 0.90920370)
(34, 0.84825188)
(35, 0.84589036)
(36, 0.81187872)
};
\addplot[blue] coordinates {
(1, 1.0224042)
(2, 1.0772120)
(3, 1.1881026)
(4, 0.81473780)
(5, 0.86347589)
(6, 0.95931312)
(7, 0.76316429)
(8, 0.77230700)
(9, 0.81444428)
(10, 1.0154658)
(11, 1.0289205)
(12, 1.0359955)
(13, 0.86624019)
(14, 0.87867591)
(15, 0.87169772)
(16, 0.84948958)
(17, 0.83735830)
(18, 0.86106537)
(19, 0.71149470)
(20, 0.88838038)
(21, 1.0648747)
(22,  0.58839171)
(23, 0.66004239)
(24, 0.72927410)
(25, 0.53973774)
(26, 0.58648021)
(27, 0.61496560)
(28, 0.74986816)
(29, 0.78287612)
(30, 0.80717411)
(31, 0.67369456)
(32, 0.67062958)
(33, 0.66278316)
(34, 0.64254453)
(35, 0.65581125)
(36, 0.65548626)
 };
  \addplot[black] coordinates {
 (1, 1)
 (2,1)
 (3, 1)
 (4, 1)
 (5, 1)
 (6, 1)
 (7,1)
 (8, 1)
 (9, 1)
 (10, 1)
 (11, 1)
 (12, 1)
 (13, 1)
 (14, 1)
 (15,1)
 (16, 1)
 (17, 1)
 (18, 1)
 (19, 1)
 (20,1)
 (21, 1)
 (22, 1)
 (23, 1)
 (24, 1)
 (25, 1)
 (26, 1)
 (27, 1)
 (28, 1)
 (29, 1)
 (30,1)
 (31, 1)
 (32, 1)
 (33, 1)
 (34, 1)
 (35, 1)
 (36, 1)
  };
  \end{axis}
  \label{precision_beta1_T40}
\end{tikzpicture} 
\end{center}
 \end{figure}

\begin{figure}[!]
 \begin{center}
 \caption{RMSE of Estimator of $\beta_2$ 
 Over 
 RMSE \\ of FD/FOD Estimator  ($T=20$).}
 \begin{tikzpicture}
\begin{axis}[myplotstyle,
xlabel = {Design No. \\
\\},
ylabel = {Relative Precision},
legend pos = north west,
legend entries = {FD, FOD,  {}},
]
\addplot[red] coordinates {
(1, 1.3099777)
(2, 1.3468721)
(3, 1.3836616)
(4, 1.1827940)
(5, 1.1997629)
(6, 1.2570192)
(7, 1.0586073)
(8, 1.0540579)
(9, 1.0644124)
(10, 1.3993448)
(11, 1.3930911)
(12, 1.4465776)
(13, 1.2569813)
(14, 1.2624323)
(15, 1.2858836)
(16, 1.0657324)
(17, 1.0682149)
(18, 1.0739130)
(19, 1.4336476)
(20, 1.5208318)
(21, 1.7756690)
(22, 1.1511138)
(23, 1.2969339)
(24, 1.3678583)
(25, 1.0525684)
(26, 1.0483052)
(27, 1.1275699)
(28, 1.5500105)
(29, 1.6673644)
(30, 1.7633617)
(31, 1.2652281)
(32, 1.3653736)
(33, 1.5077597)
(34, 1.0711499)
(35, 1.0987688)
(36, 1.1550922)
};
\addplot[blue] coordinates {
(1, 0.99559238)
(2, 0.97041943)
(3, 1.0362068)
(4, 1.0748025)
(5, 1.0107396)
(6, 1.0724309)
(7, 1.0349429)
(8, 1.0187826)
(9, 1.0072176)
(10, 1.0372685)
(11, 1.0115944)
(12, 1.0115020)
(13, 1.0859885)
(14, 1.0668626)
(15, 1.0644365)
(16, 1.0363523)
(17, 1.0331362)
(18, 1.0306748)
(19, 0.97395226)
(20, 1.1365194)
(21, 1.3635870)
(22,  1.0421333)
(23, 1.0103626)
(24, 1.0716750)
(25, 1.0545219)
(26, 1.0185199)
(27, 1.0014175)
(28, 1.0490128)
(29, 1.0695670)
(30, 1.0823665)
(31, 1.0712543)
(32, 1.0644747)
(33, 1.0747714)
(34, 1.0368879)
(35, 1.0283739)
(36, 1.0355268)
 };
  \addplot[black] coordinates {
 (1, 1)
 (2,1)
 (3, 1)
 (4, 1)
 (5, 1)
 (6, 1)
 (7,1)
 (8, 1)
 (9, 1)
 (10, 1)
 (11, 1)
 (12, 1)
 (13, 1)
 (14, 1)
 (15,1)
 (16, 1)
 (17, 1)
 (18, 1)
 (19, 1)
 (20,1)
 (21, 1)
 (22, 1)
 (23, 1)
 (24, 1)
 (25, 1)
 (26, 1)
 (27, 1)
 (28, 1)
 (29, 1)
 (30,1)
 (31, 1)
 (32, 1)
 (33, 1)
 (34, 1)
 (35, 1)
 (36, 1)
  };
  \end{axis}
  \label{precision_beta2_T20}
\end{tikzpicture} 
\end{center}
 \end{figure}

\begin{figure}[!]
\begin{center}
 \caption{RMSE of Estimator of $\beta_2$ Over RMSE \\ of FD/FOD Estimator  ($T=40$).}
 \begin{tikzpicture}
\begin{axis}[myplotstyle,
xlabel = {Design No. \\
\\},
ylabel = {Relative Precision},
legend pos = north west,
legend entries = {FD, FOD,  {}},
]
\addplot[red] coordinates {
(1, 1.3344704)
(2, 1.3513057)
(3, 1.3797582)
(4, 1.2246631)
(5, 1.2215587)
(6, 1.2841780)
(7, 1.0701834)
(8, 1.0644485)
(9, 1.0892133)
(10, 1.5713965)
(11, 1.6207401)
(12, 1.6000041)
(13, 1.2788747)
(14, 1.3155780)
(15, 1.3183983)
(16, 1.1058795)
(17, 1.0965903)
(18, 1.1176647)
(19, 1.5486003)
(20, 1.5395278)
(21, 1.6515813)
(22, 1.1926590)
(23, 1.3016050)
(24, 1.3704736)
(25, 1.0253606)
(26, 1.0385747)
(27, 1.0726490)
(28, 1.5068704)
(29, 1.6781270)
(30, 1.7788067)
(31, 1.2639958)
(32, 1.2883883)
(33, 1.3618954)
(34, 1.0040130)
(35, 1.0329205)
(36, 1.0716094)
};
\addplot[blue] coordinates {
(1, 0.94488839)
(2, 0.91580240)
(3, 0.98542569)
(4, 1.0864406)
(5, 1.0054553)
(6, 1.0574000)
(7, 1.0520624)
(8, 1.0221005)
(9, 1.0072182)
(10, 0.99206956)
(11, 0.97337732)
(12, 0.95699053)
(13, 1.0090841)
(14, 1.0022660)
(15, 0.96816100)
(16, 1.0365806)
(17, 1.0189656)
(18, 1.0142857)
(19, 1.0036880)
(20, 1.1643151)
(21, 1.4587630)
(22,  1.0630131)
(23, 1.0079871)
(24, 1.1071384)
(25, 1.0343606)
(26, 1.0104188)
(27, 0.97134822)
(28, 0.95775105)
(29, 0.95189605)
(30, 0.96713699)
(31, 0.94806477)
(32, 0.91779596)
(33, 0.89913320)
(34, 0.94230251)
(35, 0.93845261)
(36, 0.93348591)
 };
  \addplot[black] coordinates {
 (1, 1)
 (2,1)
 (3, 1)
 (4, 1)
 (5, 1)
 (6, 1)
 (7,1)
 (8, 1)
 (9, 1)
 (10, 1)
 (11, 1)
 (12, 1)
 (13, 1)
 (14, 1)
 (15,1)
 (16, 1)
 (17, 1)
 (18, 1)
 (19, 1)
 (20,1)
 (21, 1)
 (22, 1)
 (23, 1)
 (24, 1)
 (25, 1)
 (26, 1)
 (27, 1)
 (28, 1)
 (29, 1)
 (30,1)
 (31, 1)
 (32, 1)
 (33, 1)
 (34, 1)
 (35, 1)
 (36, 1)
  };
  \end{axis}
 \label{precision_beta2_T40}
\end{tikzpicture} 
\end{center}
 \end{figure}

Figures  \ref{precision_beta1_T20} through \ref{precision_beta2_T40} are line plots of relative precision estimates for the 36 designs.  A relative precision estimate is an RMSE ratio.  Specifically, an FD relative precision estimate  is an FD RMSE divided by the corresponding FD/FOD RMSE, and    a FOD relative precision estimate  is an FOD RMSE divided by the FD/FOD RMSE.  Figures \ref{precision_beta1_T20} through \ref{precision_beta2_T40} plot the FD and FOD GMM relative precision estimates for the 36 designs.  Red line plots graph the FD relative precisions, whereas blue line plots graph the FOD relative precisions. Figures \ref{precision_beta1_T20} and \ref{precision_beta1_T40} provide relative precisions for $\beta_1$ for $T=20$ and $T=40$, whereas Figures \ref{precision_beta2_T20} and \ref{precision_beta2_T40} give the same for $\beta_2$.  For all estimates, $n= 200$.
 
The plots make conclusions about relative precision obvious visually.  If an estimator's relative precision plot is above one, the estimator is less precise than the FD/FOD GMM estimator in terms of RMSE.  If the estimator's plot is below one, the estimator is more precise than the FD/FOD estimator.  

The plots  illustrate the effect of the absence, or presence, of bias on precision.  For example, although the FOD GMM estimator of $\beta_1$ has a larger variance than the corresponding FD/FOD GMM estimator,  the bias of the FOD estimator  is less---in fact, much less, especially as $T$ increases relative to $n$.  Hence, the FOD estimator of $\beta_1$ is usually more precise than the FD/FOD estimator.   On the other hand, the FD/FOD GMM estimator of $\beta_2$ generally had smaller RMSE than the FOD GMM estimator of $\beta_2$ for $T=20$.  The FOD estimator of $\beta_2$, however, became more competitive relative to the FD/FOD estimator in terms of RMSE for $T=40$. Again, finite sample bias, or its absence, provides the explanation.  The finite sample bias of the FD/FOD GMM estimator was less pronounced when estimating $\beta_2$ than it was when  $\beta_1$ was estimated, especially for $T=20$.

On the other hand, the FD GMM estimators of $\beta_1$ and $\beta_2$ have larger variances than the FD/FOD GMM estimators, but, unlike the FOD estimators, they do not benefit from having much less bias than the FD/FOD estimators. Consequently,
the FD estimators of $\beta_1$ and $\beta_2$ are less precise than the FD/FOD estimators of $\beta_1$ and $\beta_2$.  

\section{Conclusion \label{Conclusion}}

When using panel data to estimate a dynamic regression, researchers typically remove fixed effects by transforming the data.  Many transformations are available, but---due to historical precedence---differencing the data became a widely adopted approach.  However, when not all of the available instrumental variables are used, how the data are transformed matters, and differencing may not be the best transformation.  

The results provided in this paper show that transforming the data using forward orthogonal deviations  produces a GMM estimator, $\widehat{\boldsymbol{\beta}}$, such that whether or not the distribution of $\sqrt{nT}(\widehat{\boldsymbol{\beta}} - \boldsymbol{\beta})$ is centered at a vector of zeros, as $n,T \rightarrow \infty$, depends on how fast the largest number of instruments used in a period increases with $T$.   For the absence of large sample bias, this number must increase more slowly than $T^{1/2}$ increases. This observation is important because the reliability of large sample confidence intervals and test statistics based on an FOD GMM estimator depends on the distribution of $\sqrt{nT}(\widehat{\boldsymbol{\beta}} - \boldsymbol{\beta})$ being centered at a vector of zeros. 

\section*{Appendix: Proofs}

The proof of Theorem~\ref{bound_thm} draws on  Lemma~\ref{q_t_lemma}. 

\sloppy
\begin{customlemma}{A.1} \label{q_t_lemma}
	Assume $ \text{rank}\left(  \boldsymbol{Z}_t\right) = q_t   $ wp1  and define $p_{n,t} := n \, \boldsymbol{z}_{1,t}^{\prime} \left(  \boldsymbol{Z}_{t}'\boldsymbol{Z}_{t} \right) ^{-1}\boldsymbol{z}_{1,t} $. If $ \boldsymbol{z}_{i,t} $ is identically distributed across $ i $, then $ E(p_{n,t}) = q_t$. 
\end{customlemma}

\noindent{\em Proof}: By a well-known result for the trace of a projection matrix, we have $ \text{tr}\left(\boldsymbol{P}_t\right) = \text{rank}\left( \boldsymbol{Z}_t\right) $. And $\text{rank}\left( \boldsymbol{Z}_t\right) = q_t  $ wp1 by assumption.  But $ \text{tr}\left( 	 \boldsymbol{P}_t \right) = \sum_{i=1}^{n}\boldsymbol{z}_{i,t}' \left( \boldsymbol{Z}_t'\boldsymbol{Z}_t \right)^{-1} \boldsymbol{z}_{i,t}  $.
Hence, $ \sum_{i=1}^{n}E[ \boldsymbol{z}_{i,t}' \left( \boldsymbol{Z}_t'\boldsymbol{Z}_t \right)^{-1} \boldsymbol{z}_{i,t} ] = q_t$.  Therefore, if $ \boldsymbol{z}_{i,t} $ is identically distributed across $ i $, then $ n E[ \boldsymbol{z}_{1,t}' \left( \boldsymbol{Z}_t'\boldsymbol{Z}_t \right)^{-1} \boldsymbol{z}_{1,t} ] = q_t $. 

\fussy

\subsection*{Theorem \ref{bound_thm} proof}

First note that  
\begin{equation} \label{b_nT}
 \sqrt{nT} \boldsymbol{b}_{n,T} 
=  \sum_{t=1}^{T-1}c_t^2 \boldsymbol{X}_t'  \boldsymbol{P}_t \left(\boldsymbol{v}_t - \overline{\boldsymbol{v}}_t \right) - \sum_{t=1}^{T-1}c_t^2 \overline{\boldsymbol{X}}_t' \boldsymbol{P}_t \left(\boldsymbol{v}_t - \overline{\boldsymbol{v}}_t \right) .
\end{equation} 
Moreover, Assumption A1 implies $ (\boldsymbol{w}_{i,t}',v_{i,t+s}) $ is independent of $ \boldsymbol{w}_{j,t} $ for $ i \ne j $. Thus, if we set  $ \boldsymbol{w}_t' := (\boldsymbol{w}_{1,t}',\ldots, \boldsymbol{w}_{n,t}') $, then $ E\left(v_{i,t+s}|\boldsymbol{w}_{t}  \right) = E\left(v_{i,t+s}|\boldsymbol{w}_{i,t}  \right)  $.  And, A1, A3, and the law of iterated expectations imply $ E\left(v_{i,t+s}|\boldsymbol{w}_{i,t}  \right) = E\left[ E\left(v_{i,t+s}|\boldsymbol{w}_{i,t+s} \right)|\boldsymbol{w}_{i,t}\right]  = E\left( 0 | \boldsymbol{w}_{i,t}\right)  = 0$  ($ s \ge 0 $).  Hence, $  E\left(\boldsymbol{v}_t - \overline{\boldsymbol{v}}_t |\boldsymbol{w}_t\right) = \boldsymbol{0} $, and thus $ E\left[\boldsymbol{X}_t'  \boldsymbol{P}_t \left(\boldsymbol{v}_t - \overline{\boldsymbol{v}}_t \right) \right] =  E\left[\boldsymbol{X}_t'  \boldsymbol{P}_t E\left(\boldsymbol{v}_t - \overline{\boldsymbol{v}}_t |\boldsymbol{w}_t\right) \right] = \boldsymbol{0}$.  From this observation and Eq. (\ref{b_nT}), it follows that
\begin{equation} \label{E_b_nT}
		\sqrt{nT}  E(\boldsymbol{b}_{n,T})  =  - \sum_{t=1}^{T-1} \frac{1}{T-t+1}\left( 
		\boldsymbol{s}_{1,T-t} - \boldsymbol{s}_{2,T-t} 
		\right) , 
\end{equation}
where $ \boldsymbol{s}_{1,T-t} := \sum_{s=1}^{T-t}E\left( \boldsymbol{X}_{t+s}' \boldsymbol{P}_t \boldsymbol{v}_{t}\right) $ and $ \boldsymbol{s}_{2,T-t} := \sum_{s=1}^{T-t}E\left( \boldsymbol{X}_{t+s}' \boldsymbol{P}_t \overline{\boldsymbol{v}}_{t}\right) $.

\sloppy
In order to evaluate $ \boldsymbol{s}_{1,T-t}$, note that  the  $ k $th entry of $ E\left( \boldsymbol{X}_{t+s}' \boldsymbol{P}_t \boldsymbol{v}_{t}\right) $ is 
\begin{equation} \label{p_nt_step_1}
	\sum_{i=1}^n E\left( x_{i,t+s,k} \boldsymbol{z}_{i,t}'(\boldsymbol{Z}_t'\boldsymbol{Z}_t)^{-1}\boldsymbol{Z}_t'\boldsymbol{v}_{t}\right) =  \sum_{i=1}^n  E\left[   \boldsymbol{z}_{i,t}'(\boldsymbol{Z}_t'\boldsymbol{Z}_t)^{-1} \sum_{j=1}^{n} \boldsymbol{z}_{j,t} E\left(  v_{j,t} x_{i,t+s,k} | \boldsymbol{w}_t \right) \right] 
\end{equation}
And, by the law of iterated expectations, $ E\left(v_{j,t}x_{i,t+s,k} | \boldsymbol{w}_t \right) = E\left[x_{i,t+s,k}E\left( v_{j,t} | \boldsymbol{w}_t, \boldsymbol{x}_{i,t+s}  \right)  | \boldsymbol{w}_t \right]$.  Given $ (\boldsymbol{w}_{j,t}', v_{j,t}) $ is independent of $ \boldsymbol{w}_{i,T} $ for $ j \ne i $, it follows that $ E\left( v_{j,t}| \boldsymbol{w}_t, \boldsymbol{x}_{i,t+s} \right) =  E\left( v_{j,t} | \boldsymbol{w}_{j,t} \right)$ for $ j \ne i $.  Moreover,  $  E\left( v_{j,t} | \boldsymbol{w}_{j,t} \right)  = 0 $ by A1 and A3.  Hence, $ E\left(v_{j,t}x_{i,t+s,k} | \boldsymbol{w}_t \right) = 0 $ for $ j \ne i $.  On the other hand,  $ E\left( v_{j,t} x_{i,t+s,k} | \boldsymbol{w}_t \right) = E\left(  v_{i,t} x_{i,t+s,k} | \boldsymbol{w}_t \right)$ for $ j = i $. 
Hence, $ \sum_{j=1}^{n}\boldsymbol{z}_{j,t} E\left( v_{j,t} x_{i,t+s,k} | \boldsymbol{w}_t \right) = \boldsymbol{z}_{i,t} E\left(  v_{i,t} x_{i,t+s,k} | \boldsymbol{w}_t \right) $, which implies the right-hand side of (\ref{p_nt_step_1}) is 
\begin{equation} \label{lmm_4_equality}
	\sum_{i=1}^n  E\left[     \boldsymbol{z}_{i,t}'(\boldsymbol{Z}_t'\boldsymbol{Z}_t)^{-1}\boldsymbol{z}_{i,t} E\left(  v_{i,t} x_{i,t+s,k} | \boldsymbol{w}_t \right) \right] 
	= E\left[ p_{n,t}E\left( v_{1,t}x_{1,t+s,k} | \boldsymbol{w}_t  \right)\right] , 
\end{equation}
where the equality in Eq. (\ref{lmm_4_equality}) follows from the fact that the $ \boldsymbol{u}_i $s are identically distributed and  $ p_{n,t} = n \, \boldsymbol{z}_{1,t}' \left( \boldsymbol{Z}_t'\boldsymbol{Z}_t \right)^{-1} \boldsymbol{z}_{1,t} $.  Moreover,  A1 implies $ E\left(v_{1,t}x_{1,t+s,k} |\boldsymbol{w}_t\right) = E\left(v_{1,t}x_{1,t+s,k} |\boldsymbol{w}_{1,t}\right)  $.  Furthermore, from A3 and the definition of conditional covariance, we have $ \gamma_{k,t,s} =  E(v_{1,t} x_{1,t+s,k}|\boldsymbol{w}_{1,t}) $.     The preceding observations imply  the  $ k $th entry of $ E( \boldsymbol{X}_{t+s}' \boldsymbol{P}_t \boldsymbol{v}_{t}) $ is $ E( p_{n,t} \gamma_{k,t,s}) $. This conclusion and the definition of $\boldsymbol{s}_{1,T-t}$ implies the $k$th entry in  $\boldsymbol{s}_{1,T-t}$ is
$  
		 s_{1,T-t,k}  =  E(p_{n,t} \sum_{s=1}^{T-t} \gamma_{k,t,s})$ ($1 \le t \le T-1$).
Moreover,
\begin{equation} \label{s1}
		 \left|s_{1,T-t,k} \right| \le  E\left(p_{n,t} \left| \sum_{s=1}^{T-t} \gamma_{k,t,s}\right|\right) \le E\left(p_{n,t}\right) M = q_t M,
\end{equation} 
where the second inequality in (\ref{s1}) follows from A5 and the equality on the far right-hand side follows from Lemma \ref{q_t_lemma}.

Next, the entries in $\boldsymbol{s}_{2,T-t}$ are evaluated.  To that end, note that by arguments similar to those used to establish Eq. (\ref{lmm_4_equality}),  the  $ k $th entry in $ E\left( \boldsymbol{X}_{t+s}' \boldsymbol{P}_t \overline{\boldsymbol{v}}_{t}\right) $ is $ E\left( p_{n,t}\overline{v}_{1,t}x_{1,t+s,k} \right) $.
Moreover, $ E\left( p_{n,t}v_{1,t+r}x_{1,t+s,k} \right) = E\left[ p_{n,t}x_{1,t+s,k} E\left( v_{1,t+r}| \boldsymbol{w}_{t+s} \right) \right] = 0$ if $ r \geq s $. Hence, for $ s=1 $, we get $ \left( T-t\right)  E\left( p_{n,t}\overline{v}_{1,t}x_{1,t+s,k} \right) =   E( p_{n,t}x_{1,t+1,k} \sum_{r=1}^{T-t} {v}_{1,t+r} )  = 0 $. And, when $ t=T-1 $, it must be that $ s=1 $. Thus, for $ t=T-1 $ and $s=1$, we get $ \left( T-t\right)  E\left( p_{n,t}\overline{v}_{1,t}x_{1,t+s,k} \right) = E\left(p_{n,T-1}v_{1,T}x_{1,T,k}  \right) =0 $. On the other hand,  for $ 1 \le t \le T-2 $ and $ s \geq 2 $, we have
\begin{equation} \label{towards_s2}
	\begin{split}
	(T-t)  E\left(p_{n,t}\overline{v}_{1,t}x_{1,t+s,k} \right)  & =   E\left( p_{n,t} x_{1,t+s,k} \sum_{r=1}^{T-t} v_{1,t+r} \right)  \\
	& =    E\left( p_{n,t} x_{1,t+s,k} \sum_{r=1}^{s-1}v_{1,t+r} \right)  \\
	& =    E\left[    p_{n,t}  \sum_{r=1}^{s-1}E\left(v_{1,t+r}x_{1,t+r+ (s-r),k}| \boldsymbol{w}_{t+r}  \right) \right] \\
	& =    E\left[    p_{n,t}  \sum_{r=1}^{s-1}\gamma_{k,t+r,s-r}\right].
	\end{split}
\end{equation}
Moreover, for $T \ge 3$,
\begin{equation} \label{towards_s2_bnd}
\begin{split}
\sum_{s=2}^{T-t}\sum_{r=1}^{s-1}\gamma_{k,t+r,s-r} & = \underset{(r=1)}{\gamma_{k,t+1,1}} \qquad \qquad \qquad \qquad \qquad \qquad \qquad \qquad \quad (s=2) \\
& + \underset{(r=1)}{\gamma_{k,t+1,2}} + \underset{(r=2)}{\gamma_{k,t+2,1}} \qquad \qquad \qquad \qquad \qquad \qquad \quad (s=3) \\
& + \cdots \\
& + \underset{(r=1)}{\gamma_{k,t+1,T-t-1}} + \underset{(r=2)}{\gamma_{k,t+2,T-t-2}} + \cdots + \underset{(r=T-t-1)}{\gamma_{k,T-1,1}} \qquad  (s=T-t) \\
& = \sum_{r=1}^{T-t-1} \sum_{j=1}^{T-t-r} \gamma_{k,t+r,j} .
\end{split}
\end{equation}
It follows from Eq.s (\ref{towards_s2}) and (\ref{towards_s2_bnd}),  $E\left(p_{n,T-1}v_{1,T}x_{1,T,k}  \right) =0 $, and the definition of $  \boldsymbol{s}_{2,T-t} $ that the $k$th entry in  $  \boldsymbol{s}_{2,T-t} $ is $
	s_{2,T-t,k}   =  (T-t)^{-1} E(p_{n,t} \sum_{r=1}^{T-t-1} \sum_{j=1}^{T-t-r} \gamma_{k,t+r,j} )$ for  $1 \le t \le T-2 $, and 
	$s_{2,T-t,k} = 0$ for $t=T-1$.
Hence, $\left|s_{2,T-t,k}\right| = 0$ for $t=T-1$, and
\begin{equation} \label{s2}
	\left|s_{2,T-t,k}\right|   \le  \frac{1}{T-t} E\left(p_{n,t}\sum_{r=1}^{T-t-1}\left|\sum_{j=1}^{T-t-r} \gamma_{k,t+r,j}  \right| \right) 
	 \le  E\left(p_{n,t}\right) \frac{1}{T-t} \sum_{r=1}^{T-t-1}M  
	 \le q_t M
\end{equation}
for $1 \le t \le T-2$.

\fussy

Let $b_{n,T,k}$ denote the $k$th entry in $\boldsymbol{b}_{n,T}$ ($k=1,\ldots,K$). Expression (\ref{E_b_nT}),  an obvious inequality, and the inequalities in (\ref{s1}) and (\ref{s2}) imply
\begin{equation*}
\left| E\left(b_{n,T,k}\right) \right| \le \frac{1}{\sqrt{nT}} \sum_{t=1}^{T-1} \frac{1}{T-t+1}\left( \left| s_{1,T-t,k}\right| + \left| s_{2,T-t,k} \right| 
		\right) \le \frac{2}{\sqrt{nT}} M \sum_{t=1}^{T-1} \frac{q_t}{T-t+1}.
\end{equation*}
Moreover, given $q_t \le q_{T}^{\ast}$ for all $t$, we have 
 $\sum_{t=1}^{T-1} q_t/(T-t+1) \le q_{T}^{\ast} \sum_{t=1}^{T-1}1/(T-t+1)$.
And, $ \sum_{t=1}^{T-1}1/(T-t+1) = \sum_{t=1}^{T}1/t - 1$.  Furthermore, $  \sum_{t=1}^{T}1/t - 1 \le \ln T $ for $ T \ge 1 $ \citep[see,e.g.,][p. 47]{Havil2003}. The foregoing implies that, for $ T \ge 3 $,
\begin{equation} \label{bound}
\left|E\left(b_{n,T,k}\right) \right|  \le 2 M \frac{q_{T}^{\ast}}{\sqrt{nT}} \ln T   \qquad (k=1,\ldots,K).
\end{equation} 
Eq. (\ref{bound}) and $q_T^{\ast} = O(T^{\alpha})$ imply
$
 E\left(b_{n,T,k}\right)  = O\left(( T^{\alpha}\ln T)/\sqrt{nT}\right)$ for    $k=1,\ldots,K$.  The conclusion of the theorem follows from this observation and Assumption A4.

\subsection*{Lemma \ref{AR_K_lemma} proof}
Assumption A5 will be verified for $y_{1,t-1}$.  Verification of A5 for $y_{1,t-2},\ldots, y_{1,t-K}$ is similar.

\sloppy
In order to verify A5  for $y_{1,t-1}$, first note that if the roots of the characteristic equation  all lie outside the unit circle, then $y_{1,t}$ can be expressed as
\begin{equation*}
	y_{1,t} = \mu_1 + \sum_{j=0}^{\infty}\psi_{j} v_{1,t-j},
\end{equation*}
where $\mu_1 = \eta_1/\left(1 - \beta_1 - \beta_2 - \cdots - \beta_K\right)  $ and $\sum_{j=0}^{\infty}\left| \psi_{j} \right| < \infty $ \citep[see, e.g.,][pp. 58--59]{Hamilton}.  Hence, 
\begin{equation} \label{E_vy}
	\begin{split}
	E\left( v_{1,t}y_{1,t+s-1 } | \boldsymbol{w}_{1,t} \right) & = E\left(  v_{1,t} \mu_1 | \boldsymbol{w}_{1,t} \right) + \sum_{j=0}^{\infty}\psi_{j} E\left( v_{1,t}v_{1,t+s-1-j}| \boldsymbol{w}_{1,t} \right)\\
	& = \sum_{j=0}^{\infty}\psi_{j} E\left( v_{1,t}v_{1,t+s-1-j}| \boldsymbol{w}_{1,t} \right).
	\end{split}
\end{equation}
where the second equality follows from the fact that $ E\left(  v_{1,t} \mu_1 | \boldsymbol{w}_{1,t} \right) =  E\left[ \mu_1 E\left(  v_{1,t}  |\eta_1, \boldsymbol{w}_{1,t} \right)|\boldsymbol{w}_{1,t}\right] = E\left[ \mu_1 E\left(  v_{1,t}   \right)|\boldsymbol{w}_{1,t}\right] = E\left( \mu_1 0 \right) = 0 $. 

In order to evaluate $ E\left( v_{1,t}v_{1,t+s-1-j}| \boldsymbol{w}_{1,t} \right) $, recall that $ \boldsymbol{w}_{1,t}' = (y_{1,1-K},\ldots,y_{1,t-1}) $ is past information.  Moreover, if $ s \ge j+1$, then $ v_{1,t+s-1-j} $ is a current or future error, and because current and future errors are independent of past information, we have $ E\left( v_{1,t}v_{1,t+s-1-j}| \boldsymbol{w}_{1,t} \right) = E\left( v_{1,t}v_{1,t+s-1-j} \right)  $. Also, $ E\left( v_{1,t}v_{1,t+s-1-j} \right) = \sigma^2 $ or $ 0 $ according as $ j = s-1 $ or $ j \ne s-1 $. On the other hand, suppose $ s < j+1$, in which case $ v_{1,t+s-1-j} $ is a past error. To evaluate $ E\left( v_{1,t}v_{1,t+s-1-j}| \boldsymbol{w}_{1,t} \right) $ for such cases, let ${\cal{F}}_{1,t-1}$ be the $\sigma$-field generated by $\boldsymbol{w}_{1,t} = (y_{1,1-K},\ldots,y_{1,t-1})'$. Then $E\left( v_{1,t}v_{1,t+s-1-j}| \boldsymbol{w}_{1,t} \right) = E\left( v_{1,t}v_{1,t+s-1-j}| {\cal{F}}_{1,t-1} \right)$. Next, let ${\cal{F}}_{t-1}$ denote the $\sigma$-field generated by $\eta_1$ and the  past $ y_{1,t}$s, i.e., ${\cal{F}}_{t-1} := \sigma\left(\ldots,  y_{1,t-2},y_{1,t-1},\eta_1 \right) $.   By the law of iterated expectations, $E\left( v_{1,t}v_{1,t+s-1-j}| {\cal{F}}_{1,t-1} \right) = E\left[ E\left( v_{1,t}v_{1,t+s-1-j}| {\cal{F}}_{t-1}  \right)|{\cal{F}}_{1,t-1} \right] $.  Moreover, given we are now considering the cases for which $ s-1-j = -1, -2, \ldots $,  we get $E\left( v_{1,t}v_{1,t+s-1-j}| {\cal{F}}_{t-1}  \right) = \left( \beta(L)y_{1,t+s-1-j} - \eta_1\right) E\left( v_{1,t}| {\cal{F}}_{t-1}  \right) = \left( \beta(L)y_{1,t+s-1-j} - \eta_1\right) E\left( v_{1,t}  \right) = 0$. Hence, $ E\left( v_{1,t}v_{1,t+s-1-j}| \boldsymbol{w}_{1,t} \right) =0$ for $ s < j+1$. The preceding shows $ E\left( v_{1,t}v_{1,t+s-1-j}| \boldsymbol{w}_{1,t} \right) = \sigma^2$ or $ 0 $ according as $ j = s-1 $ or $ j \ne s-1 $.  

The conclusion of the last paragraph and (\ref{E_vy}) imply $ 	E\left( v_{1,t}y_{1,t+s-1} | \boldsymbol{w}_{1,t} \right) = \sigma^2 \psi_{s-1} $.  Moreover, because $E\left( v_{1,t}| \boldsymbol{w}_{1,t} \right) = 0$, we have $\text{cov}\left( v_{1,t},y_{1,t+s-1 } | \boldsymbol{w}_{1,t} \right)  = E\left( v_{1,t}y_{1,t+s-1 } | \boldsymbol{w}_{1,t} \right)$. Hence, $\text{cov}\left( v_{1,t},y_{1,t+s-1} | \boldsymbol{w}_{1,t} \right)  =  \sigma^2 \psi_{s-1} $ for $s \ge 1$. 
And $\sum_{j=0}^{\infty}\left| \psi_{j} \right| < \infty $ by assumption. Therefore, Assumption A5 is satisfied for the first regressor $ y_{1,t-1} $.  Similar arguments  verify A5 for  $ y_{1,t-2},\ldots,y_{1,t-K}  $.

\subsection*{Theorem \ref{normality_thm} proof}

Note that  $\boldsymbol{\Pi}_t = \boldsymbol{Q}_t^{-1} \boldsymbol{C}_t$ are the linear projection parameters in the $K$ reduced form equations
$
\ddot{\boldsymbol{x}}_{i,t} = \boldsymbol{\Pi}_t' \boldsymbol{z}_{i,t} + \boldsymbol{r}_{i,t},
$
where $\boldsymbol{r}_{i,t}' = (r_{i,t,1},\ldots,r_{i,t,K})$ are  reduced form errors. Hence, $
\ddot{\boldsymbol{X}}_{i,t} = \boldsymbol{Z}_{t}\boldsymbol{\Pi}_t  + \boldsymbol{R}_{t}$, where $\boldsymbol{R}_{t}' = (\boldsymbol{r}_{1,t},\ldots, \boldsymbol{r}_{n,t})$.  Therefore, $\boldsymbol{b}_{n,T} = (1/\sqrt{nT})\sum_{t=1}^{T-1}\boldsymbol{\Pi}_t' \boldsymbol{Z}_{t}'\ddot{\boldsymbol{v}}_{t} + (1/\sqrt{nT})\sum_{t=1}^{T-1}\boldsymbol{R}_t' \boldsymbol{P}_{t}\ddot{\boldsymbol{v}}_{t}$.  Moreover, $(1/\sqrt{nT})\sum_{t=1}^{T-1}\boldsymbol{R}_t' \boldsymbol{P}_{t}\ddot{\boldsymbol{v}}_{t} = (1/\sqrt{T})\sum_{t=1}^{T-1}(1/n)\boldsymbol{R}_t' \boldsymbol{Z}_{t}\left[(1/n)\boldsymbol{Z}_{t}'\boldsymbol{Z}_{t}\right]^{-1}(1/\sqrt{n})\boldsymbol{Z}_{t}'\ddot{\boldsymbol{v}}_{t}$.  Because $E(\boldsymbol{r}_{i,t}\boldsymbol{z}_{i,t}') = E(\ddot{\boldsymbol{x}}_{i,t}\boldsymbol{z}_{i,t}') - \boldsymbol{\Pi}_t' E(\boldsymbol{z}_{i,t}\boldsymbol{z}_{i,t}') = \boldsymbol{C}_t' - \boldsymbol{C}_t'\boldsymbol{Q}_t^{-1}\boldsymbol{Q}_t = \boldsymbol{0}$, we have that $(1/n)\boldsymbol{R}_t' \boldsymbol{Z}_{t} = (1/n)\sum_{i=1}^n\boldsymbol{r}_{i,t} \boldsymbol{z}_{i,t}' \overset{p}\rightarrow \boldsymbol{0}$, as $n \rightarrow \infty$, by the law of large numbers.  Similarly, $(1/n)\boldsymbol{Z}_{t}'\boldsymbol{Z}_{t} = (1/n)\sum_{i=1}^n\boldsymbol{z}_{i,t}\boldsymbol{z}_{i,t}' \overset{p}\rightarrow \boldsymbol{Q}_t$ as $n \rightarrow \infty$.  Moreover, Assumption A3 implies $ E(\boldsymbol{z}_{1,t}\ddot{v}_{1,t})= \boldsymbol{0} $, and Assumption A6 implies that  $ \text{Var}(\boldsymbol{z}_{1,t}\ddot{v}_{1,t})$ has finite entries. Therefore, by the central limit theorem for i.i.d. random vectors, we have $ (1/\sqrt{n}) \boldsymbol{Z}_{t}'\ddot{\boldsymbol{v}}_{t} \overset{d}{\rightarrow} N(\boldsymbol{0}, \text{Var}(\boldsymbol{z}_{1,t}\ddot{v}_{1,t}))$ as $ n \rightarrow \infty $. Hence,  $  (1/\sqrt{n})\boldsymbol{Z}_{t}'\ddot{\boldsymbol{v}}_{t} = O_p(1) $ for each $ t $.  It follows  that 
$
		\boldsymbol{b}_{n,T} - (1/\sqrt{nT})\sum_{t=1}^{T-1}\boldsymbol{\Pi}_t' \boldsymbol{Z}_{t}'\ddot{\boldsymbol{v}}_{t}    
 \overset{p}{\rightarrow} \boldsymbol{0}$ as $n \rightarrow \infty$.
Therefore,  the asymptotic distribution of $  \boldsymbol{b}_{n,T} $, as  $ n \rightarrow \infty $, is the same as the asymptotic distribution of $  (1/\sqrt{nT})\sum_{t=1}^{T-1} \boldsymbol{\Pi}_t' \boldsymbol{Z}_{t}'\ddot{\boldsymbol{v}}_{t}$, as $ n \rightarrow \infty $.

To evaluate the latter,  set $\boldsymbol{\epsilon}_{i,T} := (1/\sqrt{T})\sum_{t=1}^{T-1}\boldsymbol{\Pi}_t^{\prime}\boldsymbol{z}_{i,t}\ddot{v}_{i,t} $.  Note that $E(\boldsymbol{\epsilon}_{1,T}) = \boldsymbol{0}$ and $\text{Var}(\boldsymbol{\epsilon}_{1,T}) = \boldsymbol{\Omega}_T$.  By the central limit theorem for i.i.d. random vectors, we have that $(1/\sqrt{n})\sum_{i=1}^n\boldsymbol{\epsilon}_{i,T} \overset{d}\rightarrow  N(\boldsymbol{0}, \boldsymbol{\Omega}_T)$  ($n \rightarrow \infty$).  But $(1/\sqrt{n})\sum_{i=1}^n\boldsymbol{\epsilon}_{i,T} = (1/\sqrt{nT})\sum_{t=1}^{T-1} \boldsymbol{\Pi}_t^{\prime} \boldsymbol{Z}_{t}'\ddot{\boldsymbol{v}}_{t}$, and recall from the last paragraph that $\boldsymbol{b}_{n,T} - (1/\sqrt{nT})\sum_{t=1}^{T-1}\boldsymbol{\Pi}_t^{\prime} \boldsymbol{Z}_{t}'\ddot{\boldsymbol{v}}_{t} \overset{p}\rightarrow \boldsymbol{0} $ ($n \rightarrow \infty$). Hence, $\boldsymbol{b}_{n,T} \overset{d}\rightarrow \boldsymbol{b}_T \sim N(\boldsymbol{0}, \boldsymbol{\Omega}_{T})$ ($n \rightarrow \infty$).

\fussy
The characteristic function of $\boldsymbol{b}_T$ is
$
\phi_{T}(\boldsymbol{\lambda}) := \exp\left( -(1/2)\boldsymbol{\lambda}'\boldsymbol{\Omega}_{T}\boldsymbol{\lambda}\right)
$.
Moreover, Assumption A7 implies
$
\lim_{T \rightarrow \infty}\phi_{T}(\boldsymbol{\lambda}) = \exp\left( -(1/2)\boldsymbol{\lambda}'\boldsymbol{\Omega}\boldsymbol{\lambda}\right).
$
The latter limit is the characteristic function of a multivariate normal vector with mean  $\boldsymbol{0}$ and variance-covariance matrix $\boldsymbol{\Omega}$.  Hence, $\boldsymbol{b}_{T} \overset{d}\rightarrow \boldsymbol{b} \sim N(\boldsymbol{0}, \boldsymbol{\Omega})$ ($T \rightarrow \infty$).

\sloppy
The preceding verifies $ \boldsymbol{b}_{n,T} \overset{d}{\rightarrow} \boldsymbol{b} \sim N(\boldsymbol{0}, \boldsymbol{\Omega} )$ $ ( n, T \rightarrow \infty)_{\text{seq}} $.  We have that  $\boldsymbol{A}_{n,T} \overset{p}\rightarrow \boldsymbol{A}_{T} := (1/T)  \sum_{t=1}^{T-1} \boldsymbol{C}_t ' \boldsymbol{Q}_t^{-1} \boldsymbol{C}_t $ ($n \rightarrow \infty$) by the law of large numbers, and $ \boldsymbol{A}_{T} \rightarrow \boldsymbol{A} >0 $  ($T \rightarrow \infty $) by Assumption A8. Hence,   $\sqrt{nT}(\widehat{\boldsymbol{\beta}}-\boldsymbol{\beta}) =  \boldsymbol{A}_{n,T}^{-1}\boldsymbol{b}_{n,T} \overset{d}{\rightarrow} N(\boldsymbol{0}, \boldsymbol{A}^{-1} \boldsymbol{\Omega}\boldsymbol{A}^{-1})$ $ ( n, T \rightarrow \infty)_{\text{seq}} $.

\fussy

\end{document}